\documentclass[%
 amsmath,amssymb,
]{revtex4-2}

\usepackage{dcolumn} 
\usepackage{bm}      
\usepackage{setspace}
\usepackage{amsfonts}
\usepackage{braket}

\usepackage{graphicx}

\newcommand{\dev}[3]{\frac{\text{d}^{#3} #1}{\text{d}#2^{#3}}}

\newcommand{\ff}{{\mathbf{f}}}
\newcommand{\ww}{{\mathbf{w}}}
\newcommand{\xx}{{\mathbf{x}}}

\newcommand{\zz}{{\mathbf{z}}}

\newcommand{\GG}{{\mathbf{G}}}

\newcommand{\Rset}{{\mathbb{R}}}
\newcommand{\Cset}{{\mathbb{C}}}

\usepackage{amsmath}	
\begin{document}


\title{Expanding detection bandwidth via a photonic reservoir for ultrafast optical sensing}

\author{
Yuito Ito$^{1}$,
Tomoaki Niiyama$^{2}$,
Tetsuya Asai$^{3}$,
Gouhei Tanaka$^{4,5}$,
Atsushi Uchida$^{6}$,
and 
Satoshi Sunada$^{2}$}

\affiliation{${}^{1}$Graduate School of Natural Science and Technology, Kanazawa University, Kakuma-machi, Kanazawa, Ishikawa, 920-1192, Japan\\ ${}^{2}$Faculty of Mechanical Engineering, Institute of Science and Engineering, Kanazawa University, Kakuma-machi, Kanazawa, Ishikawa, 920-1192, Japan\\ ${}^{3}$Faculty of Information Science and Technology, Hokkaido University,
Kita 14, Nishi 9, Kita-ku, Sapporo, Hokkaido, 060-0814, Japan\\ ${}^{4}$Graduate School of Engineering, Nagoya Institute of Technology, Showa-ku, Nagoya, Aichi, 466-8555, Japan\\ ${}^{5}$International Research Center for Neurointelligence, Institutes for Advanced Study, The University of Tokyo, Tokyo, 113-0033, Japan\\ ${}^{6}$Department of Information and Computer Sciences, Saitama University, 255 Shimo-Okubo, Sakura-ku, Saitama City, Saitama, 338-8570, Japan}%


\begin{abstract}
The detection of ultrafast optical and radio-frequency (RF) signals is crucial for applications ranging from high-speed communications to advanced sensing. However, conventional detectors are fundamentally constrained by their intrinsic bandwidth, limiting accurate broadband signal measurement. Here, we show that a neuromorphic photonic processing approach can overcome this limitation, enabling accurate broadband signal detection beyond the detector bandwidth. The key idea lies in the spatiotemporal encoding of input waveforms within a photonic reservoir network, which reconstructs high-frequency components otherwise inaccessible to individual detectors. We experimentally demonstrate the detection of high-speed optical phase signals with more than an eightfold effective bandwidth expansion using an on-chip silicon photonic reservoir. This approach provides a scalable and integrable platform for high-speed optical and RF signal processing, opening new opportunities in ultrafast photonics and next-generation communication systems.
\end{abstract}

\maketitle
\section{Introduction}
High-speed optical detection and sensing are essential in modern optical technologies, including optical communication \cite{Mueller:2010aa,Liu:25,Zhao_2017,photonics8010014}, high-speed imaging \cite{Goda:2009aa,Gibson:20}, millimeter/terahertz wave generation \cite{app14083410,Koch:2023aa}, photonic computing \cite{Chen:2023wi,Freire:23}, and quantum information processing \cite{Kawasaki:2024aa}. The increasing demand for higher data rates and faster response times has driven extensive research on optical sensing technologies with continually increasing bandwidths \cite{Marpaung:2019aa,10.1063/5.0146912}.

Despite these advances, conventional high-speed photodetectors still face fundamental and practical limitations that restrict their performance in many applications \cite{Han:2025_LPR,Sorger:20,Zhu:25,doi:10.1126/sciadv.abq0187,Chee19}. First, the well-known bandwidth--responsivity trade-off persists. For example, increasing the bandwidth typically requires reducing the active area or junction capacitance, which diminishes quantum efficiency and lowers the signal-to-noise ratio (SNR). Second, achieving high-speed operation often entails increased system complexity and cost, owing to the need for broadband amplifiers, impedance-matched circuits, and low-noise electronics. Whereas a variety of promising photodetectors and photodiode architectures have been reported \cite{Sorger:20,Chee19,Song:14,775466,180577,1075419,Gao:2017aa}, these limitations underscore the difficulty of simultaneously achieving ultrafast response and high sensitivity using conventional designs.

As relevant research, photonic-assisted high-speed data acquisition techniques have been explored, including photonic interleaved sampling \cite{Li:22,Yang:16,Xu_2016}, photonic time-stretch technologies \cite{photonics10070817,Godin31122022,10.1063/1.4941050,Mahjoubfar:2017aa,775471}, and photonic compressed sensing \cite{Zhou:22,Yang:23}. However, most of these techniques have been developed primarily for the sampling of high-speed radio-frequency (RF) signals rather than optical signals. Consequently, photodetection or optoelectronic conversion remains necessary for direct ultrafast optical signal measurement.

Herein, we propose an optoelectronic conversion approach for ultrafast optical sensing, inspired by a recent neuromorphic computing paradigm, namely, reservoir computing \cite{Jaeger78,Tanaka2019,Yan:2024aa}. In this framework, a dynamical system with rich internal states (the ``reservoir'') transforms input signals into a high-dimensional representation, in which simple readout operations can efficiently solve complex tasks. Photonic reservoir computing implements this concept in the optical domain, typically by leveraging the inherent dynamics of optical systems with recurrent or random network topologies. Owing to its ability to perform high-dimensional mappings, photonic reservoir computing has emerged as a powerful tool for ultrafast time-series processing \cite{VanderSandeBrunnerSoriano+2017+561+576,Brunner2013,Takano2018,Vandoorne2014,Wang2025,DWang2024,Sunada2021}.

A key idea of our approach is the use of a photonic reservoir network for spatiotemporal mapping, which distributes high-speed temporal information into multiple low-frequency channels. This transformation enables conventional narrowband photodetectors to capture broadband optical signals well beyond their intrinsic bandwidth. The main advantages of the proposed approach can be summarized as follows:
(i) Compatibility with low-cost, off-the-shelf detectors;
(ii) Applicability to non-repetitive, single-shot measurements;
(iii) Capability to detect both optical intensity and phase dynamics; and
(iv) Potential for seamless integration into silicon photonic platforms.
Additionally, in contrast to conventional optical measurement methods \cite{Roussel:2022aa,Tikan:2018aa,Suret:2016aa}, the proposed approach is free from pulse-laser constraints; therefore, the record length of the measured signal is not limited by the laser pulse duration.

Through experiments, we demonstrate broadband optical signal detection with an effective bandwidth nearly an order of magnitude larger than the detector bandwidth, using a silicon photonic reservoir chip. The findings of this study establish a new path toward bandwidth-agnostic optical sensing, bridging the gap between ultrafast optical phenomena and practical measurement hardware.

\section{Reservoir sensing: Framework}
\begin{figure}[htbp]
\centering
\includegraphics[bb = 0 0 583 441, width=13cm]{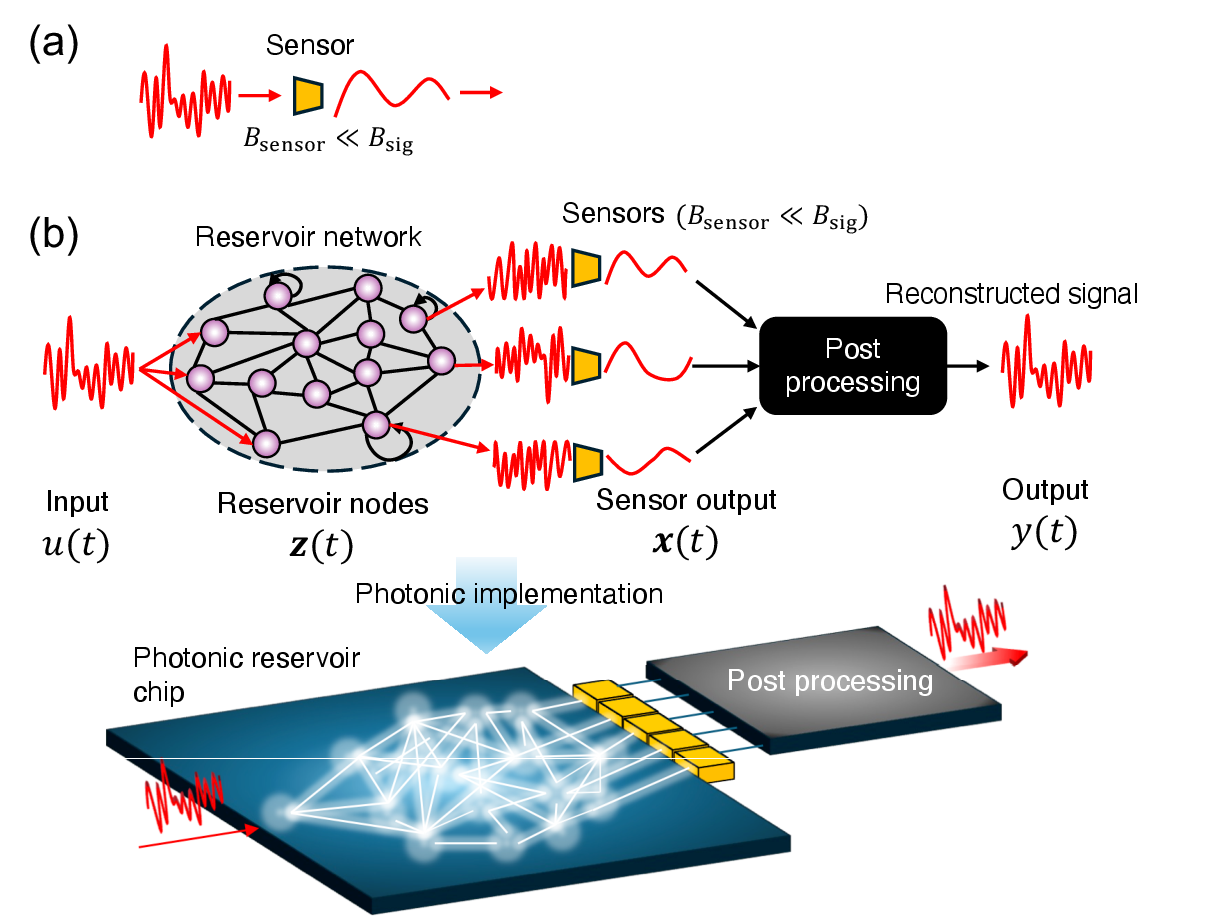}
\caption{(a) High-speed signal detection using a single sensor whose bandwidth $B_{\rm sensor}$ is narrower than the signal bandwidth $B_{\rm sig}$. (b) Conceptual schematic of the proposed sensing approach and its photonic implementation. PDs: photodetectors.}
\label{fig_concept}
\end{figure}

Figure~\ref{fig_concept} shows a conceptual schematic of the proposed sensing approach. The sensing system consists of a so-called reservoir network, e.g., a recurrent network with random connections, and multiple sensors for reservoir output detection. Herein, we assume that the sensor bandwidth $B_{\rm sensor}$ is much narrower than the signal's maximum frequency $B_{\rm sig}$ [Fig.~\ref{fig_concept}(a)], i.e.,
\begin{align}
B_{\rm sensor} \ll B_{\rm sig}.  
\end{align}
The goal of the proposed approach is to accurately capture a broadband, high-speed signal beyond the sensor bandwidth limitation. Accordingly, the high-frequency information attenuated due to the limited sensor bandwidth must be recovered. To address this issue, high-speed temporal information is encoded as spatially distributed narrow-band signals through a reservoir network. Then, the spatially distributed narrow-band signals are simultaneously measured using multiple narrow-band sensors [Fig.~\ref{fig_concept}(b)].

Let the signal to be measured (target signal) be $u(t) \in \Rset$ at time $t$. The resevoir network has $N_r$ nodes, which are represented as $\zz(t) = (z_1,z_2,\cdots,z_{N_r})^{T} \in \Cset^{N_r}$ and are governed by
\begin{align}
 \dev{\zz(t)}{t}{} = \GG(\zz(t),u(t)),
\end{align}
where $\GG(\cdot)$ represents a function characterizing the dynamical property of the reservoir network.
A subset of the reservoir nodes is measured by $N$ sensors. The sensor outputs are represented as $\xx(t) = \ff(\zz(t)) \in \mathbb{R}^N$, where $\ff(\cdot)$ and $N$ denote a certain function and the number of sensors, respectively. For instance, if the sensors act as a low-pass filter (LPF) for the intensity, the function is expressed as 
$\ff(\zz) = \mbox{LPF}(\left|\zz(t)\right|^2), $
where $\mbox{LPF}$ represents an LPF function. 

Suppose that $\xx(t)$ are sampled at a sampling time interval $\Delta t$. The target signal at time $t_j = j\Delta t,$ ($j \in \mathbb{N}$), can be inferred from sampled signal $\xx(t_j)$ via an appropriate regression model. While various models, including neural networks, can be used for this purpose, in this study, we simply used a linear model because of its low training computational cost. For the linear model, the output signal is simply expressed with a linear combination of the sampled signal sequence $\{\xx(j\Delta t)\}_{j=1}^{N_T}$ and the weight vectors as follows:
\begin{align}
y(j\Delta t) = \sum_{k=0}^{K-1}\ww_{k}^{T}\xx((j-k)\Delta t),
\label{eq_outout} 
\end{align}
where a time-multiplexing technique, which is commonly used in reservoir computing, is used to improve the reconstruction performance \cite{Sunada:21}. $K$ denotes the number of time-multiplexed steps. $\ww_{k} \in \Rset^{N}$ represents the weight vector of the $k$-th delay step, which are determined through training. 

In the training phase, weight vectors $\{\ww_{k}\}_{k=0}^{K-1}$ are optimized using multiple waveforms, $\{u_1(t_j), u_2(t_j), \cdots, u_L(t_j)\}_{j=1}^{N_T}$, which were sampled at $t_j = j\Delta t$ from $\Delta t$ to $T = N_T\Delta t$. We chose waveform $u_l(t)$ such that it has multiple frequencies in a range up to $B_{\rm sig}$ and satisfies $\langle u_l(t)u_{l'}(t)\rangle/\alpha_{ll'} = \delta_{ll'}$, where $\alpha_{ll'}$ and $\delta_{ll'}$ denote the normalization factor and Kronecker delta, respectively. See Supplementary Section~1 for more details. Then, optimal weight vectors $\{\ww^*_{k}\}_{k=0}^{K-1}$ are determined to minimize the following mean squared error (MSE):
\begin{align}
\{\ww^*_{k} 
\}_{k=0}^{K-1}
= \mbox{argmin}_{\{\ww_{k}\}}\left(
\sum_{l=1}^{L}\sum_{j=1}^{N_T}
\left|
y_l(j\Delta t) - u_l(j\Delta t)
\right|^2 
+ R\left(
\{\ww_k\}_{k=0}^{K-1}
\right)
\right), \label{eq_optimalweight}
\end{align}
where $y_l(t)$ is the output signal for the reservoir responding to input $u_l(t)$ and $R(\cdot)$ is a regularization term. Eq.~(\ref{eq_optimalweight}) can be solved via the linear regression for $R = 0$ or the ridge regression method for $R \propto \sum_k|\ww_k|^2$. 

\section{Numerical results}
To numerically validate the effectiveness of the proposed sensing approach, we employed a simple reservoir network model, known as the Echo State Network (ESN) model \cite{article}. The ESN model was trained with $L$ waveforms $\{u_l(t_j)\}_{j,l=1}^{N_T,L}$ (where $N_T = 16384$; $L = 10$; $\Delta t = 1/N_T$). In this study, the sensors were assumed to act as third-order Butterworth LPFs. Then, we define the bandwidth ratio $B$ between the bandwidth of the target (test) signal $u_{\rm test}(t)$ and that of the sensors used for signal detection as
$B = B_{\rm sig}/B_{\rm sensor}$.
As we are interested in the sensing capability for unknown signals without any prior knowledge, we used test signal $u_{\rm test}(t)$, which is linearly independent of the training signals, i.e., $\langle u_{\rm test}(t)u_l(t)\rangle = 0$. The reconstruction error was quantified using the normalized mean squared error (NMSE), which is defined as
\begin{align}
\mathrm{NMSE} = \frac{\sum_{j=1}^{N_{\rm test}} \left( y(t_j) - u_{\rm test}(t_j) \right)^{2}}{\sigma_u^{2}},
\end{align}
where $\sigma_u$ is the standard deviation of target signal $u(t)$.

\begin{figure}[htbp]
\centering
\includegraphics[bb = 0 0 565 551, width=12cm]{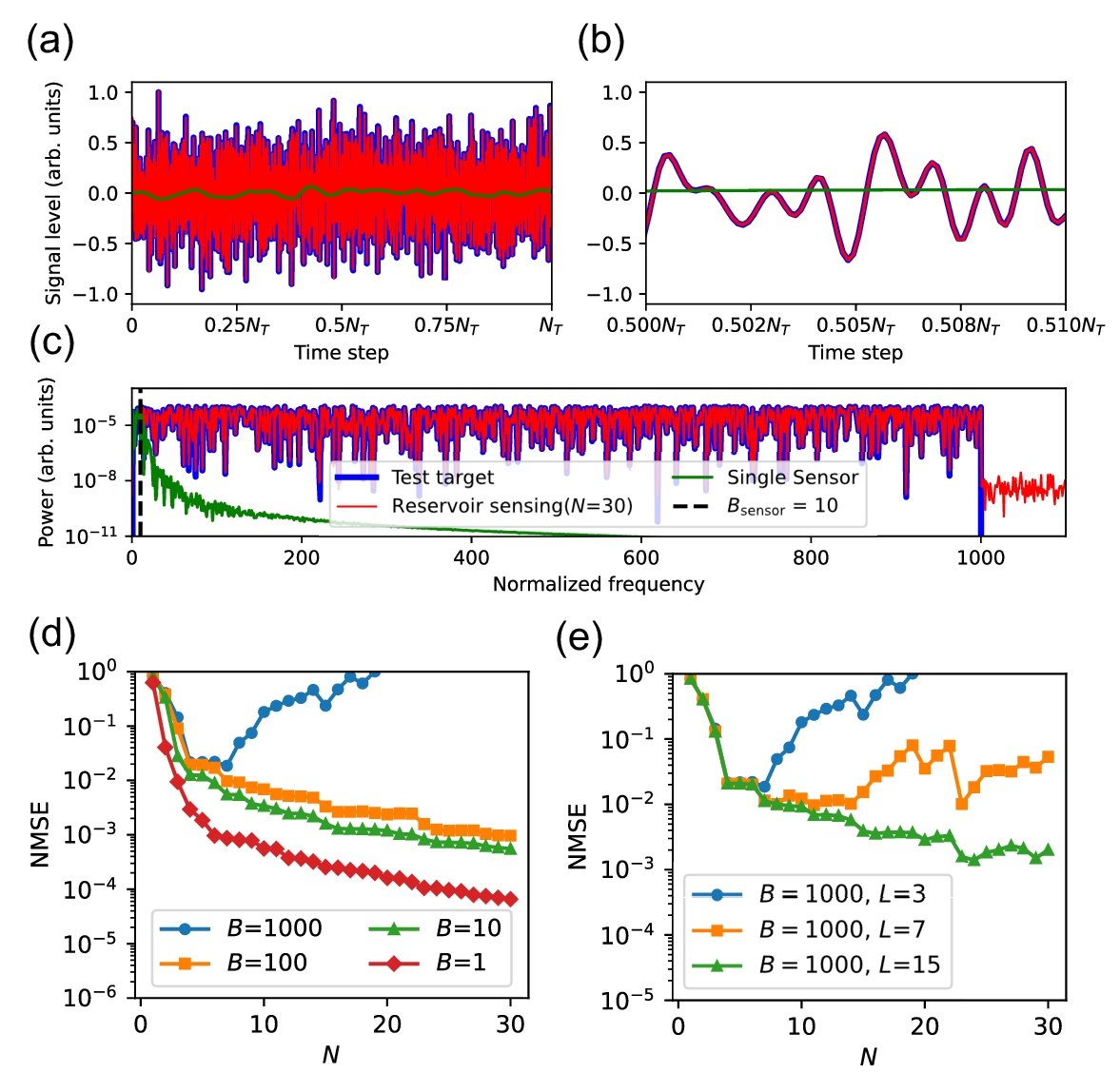}
\caption{
Numerical reconstruction results. 
(a) Original and reconstructed signals in the time domain.
(b) Enlarged view of (a).
(c) Power spectra of the original, sensor output, and reconstructed signals.
(d) NMSE versus $N$ for different bandwidth ratios $B = B_{\rm sig}/B_{\rm sensor}$ for $L=3$.
(e) NMSE versus $N$ for different training sample sizes $L$ for $B = 1000$.}
\label{fig_random}
\end{figure}

Figure~\ref{fig_random} shows the reconstruction results for broadband test signal $u_{\rm test}(t)$. In this numerical experiment, we used $N_r = 100$ and $N = 30$. The dimensionless frequency of the test signal ranges from 0 to 1,000 (Fig.~\ref{fig_random}(c)), but the high-frequency components are drastically attenuated after passing through the sensor LPF with $B_{\rm sensor} = 100$. By contrast, our reservoir-based sensing approach can recover the high-frequency components (Fig.~\ref{fig_random}(c)) and successfully reproduces the original waveform (Figs.~\ref{fig_random}(a) and (b)). The NMSE was $3.0\times10^{-4}$.

The reconstruction quality depends on the number of sensors $N$, the bandwidth ratio $B$, and the number of training data samples $L$. Figure~\ref{fig_random}(d) shows the NMSE as a function of $N$ for $B =$ 1, 10, 100, and 1000 and $L = 3$. Despite the limited number of training samples, the NMSE for $B \leq 100$ decreases monotonically as $N$ increases, confirming that the proposed approach enhances the reconstruction capability. By contrast, the NMSE for $B = 1000$ tends to deteriorate with increasing $N$. This deterioration can be mitigated by increasing $L$ (Fig.~\ref{fig_random}(e)), thereby achieving scalable NMSE reduction with respect to $N$.

\begin{figure}[htbp]
\centering
\includegraphics[bb=0 0 283 183,width=7cm]{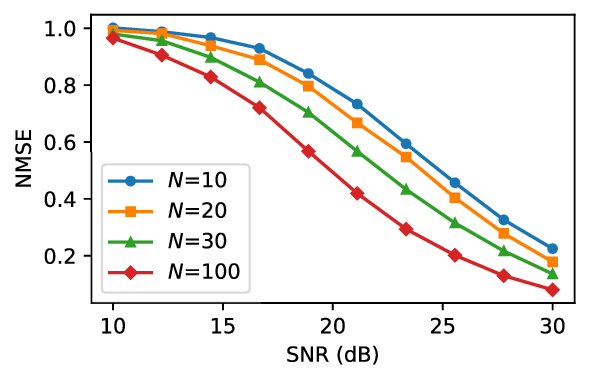}
\caption{NMSE as a function of the SNR.}
\label{fig_noise}
\end{figure}

A remarkable feature of the linear representation (Eq.~(\ref{eq_outout})) for reconstruction is its ability to reduce the effect of sensor noise. To demonstrate this feature, we assumed that each sensor output is disturbed as $x_j(t) + n_j(t)$, where $n_j(t)$ is modeled as Gaussian noise with mean $0$ and standard deviation $\sigma$, satisfying $\langle n_i(t) n_j(t') \rangle = \sigma^2\delta_{ij}\delta(t-t')$. The SNR is defined as
$\mbox{SNR} = 10 \log_{10} S/\sigma$,
where $S$ denotes the standard deviation of the target signal waveform. Figure \ref{fig_noise} shows the NMSE as a function of the SNR for various $N$ values for $B = 10$. The reconstruction quality degradation can be compensated by increasing $N$. The NMSE was roughly estimated as
$\mathrm{NMSE} \propto N^{-1/2}\exp(-\mbox{SNR})$ for $N \gg 1$
based on the central limit theorem.

\section{Experiments}
\subsection{Photonic reservoir network}
\begin{figure}[htbp]
\centering
\includegraphics[bb=0 0 511 175, width=12cm]{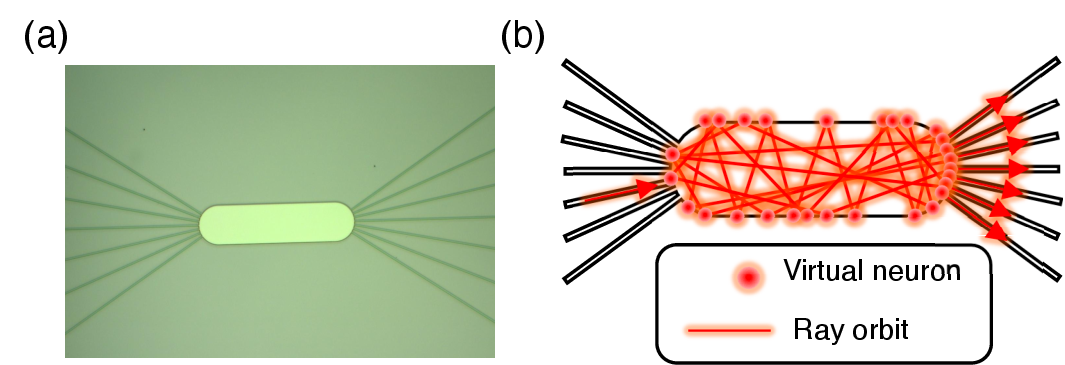}
\caption{(a)Photograph of the silicon photonic reservoir chip. (b)Schematic of a virtual reservoir network.}
\label{fig_reservoir}
\end{figure}

To experimentally validate the proposed sensing approach, we used a silicon photonic reservoir chip based on a stadium-shaped microcavity coupled to 14 single-mode waveguide channels (Fig.~\ref{fig_reservoir}(a)). The stadium-shaped microcavity is inspired by the Bunimovich stadium, which naturally supports nonlinear ray dynamics and wave chaos \cite{Bunimovich:1974aa}. The stadium-shaped cavity can inherently induce complex dynamics, resulting in efficient high-dimensional feature mapping suitable for reservoir computing. The optical signal is input to the microcavity via a single-mode waveguide channel and mixed due to the ray-chaotic multiple reflections inside the cavity, forming a complex optical network (Fig.~\ref{fig_reservoir}(b)). The cavity size is within 50 $\times$ 200 $\mu$m. The memory length for storing past information was roughly estimated as 250 ps \cite{Yamaguchi:2023aa}. Previous studies have used the microcavity reservoir for information processing, such as chaotic time-series prediction, vowel recognition, and high-speed image processing \cite{Yamaguchi:2023aa}. In this study, the microcavity reservoir was used for sensing. 

Another particular feature of the photonic microcavity-based reservoir is its capability of transforming an {\it optical temporal phase} signal, e.g., $e^{iu(t)}$, into intensity signals due to the optical interference inside the microcavity \cite{You:25}. The nonlinear phase-to-intensity transformation enables broadband temporal phase signal detection, which is inaccessible when using a single photodetector, as demonstrated next.

\subsection{Experimental setup}
\begin{figure}[htbp]
\centering
\includegraphics[bb=0 0 684 552,width=14cm]{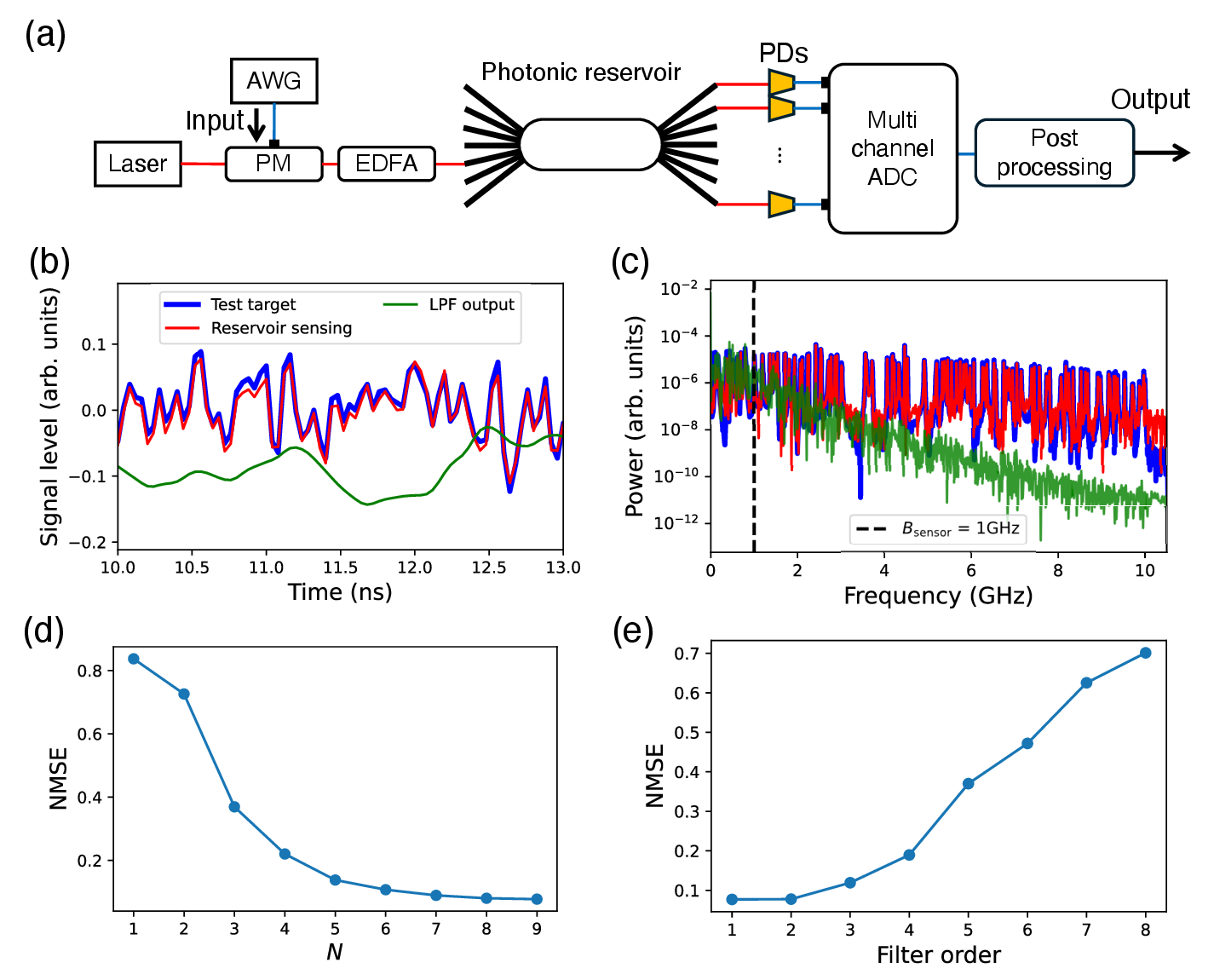}
\caption{Experimental bandwidth expansion demonstration. 
(a) Schematic of experimental setup. AWG: Arbitrary Waveform Generator; PM: Phase Modulator; EDFA: Erbium-Doped Fiber Amplifier; PD: Photodetector; ADC: Analog-Digital Converter.
(b) Time-domain comparison between the original (target), filtered, and reconstructed signals for a 10-GHz test signal with a 1-GHz filter. Bandwidth ratio $B = 10$. 
(c) Power spectra.
(d) NMSE versus $N$. 
(e) NMSE versus filter order.}
\label{fig_exp1}
\end{figure}

The experimental setup is illustrated in Fig.~\ref{fig_exp1}(a). A narrow-linewidth laser (Alnair Labs TLG-220; linewidth: 100 kHz; output power: 20 mW) was employed as the light source. The input signal $u(t)$ was generated by an arbitrary waveform generator (Tektronix AWG70002A) operating at 25 Gigasamples per second (GS/s). $u(t)$ was encoded in the optical phase using a lithium niobate phase modulator (EO Space PM-5S5-20-PFA-PFA-UV-UL; bandwidth: 16 GHz). The modulated light was injected into the photonic reservoir chip through an input waveguide channel, and the outputs from seven channels were measured using photodetectors. The detected signals were acquired with a sampling interval of $\Delta t = 0.02$ ns using a digital oscilloscope (Tektronix DPO72504DX, 50 GS/s), which was used as a multichannel analog-to-digital converter (ADC), and were processed for signal reconstruction on a personal computer. For training, $L = 10$ random signals $\{u_l(t)\}_{l=1}^{L}$ of length $N_T = 16,384$ ($T = 655.36$ ns) and maximum frequency $B_{sig} =$ 10 GHz were employed. 

\subsection{Bandwidth expansion}
To gain systematic insights on the bandwidth expansion of the proposed approach, we used InGaAs PIN photodetectors (Thorlabs RXM25BF) and emulated the narrowband detection response using a second-order Butterworth filter (cutoff frequency: 1 GHz). 
Figure~\ref{fig_exp1}(b) shows the reconstruction result for a test signal with a bandwidth of $B_{\rm sig} = $10 GHz, which is linearly uncorrelated with all training signals. The filtered reservoir outputs were attenuated in the high-frequency region; therefore, the power spectrum was completely different from that of the original test signal. Meanwhile, the proposed reservoir sensing approach recovered the high-frequency information, as shown by the red curve in Fig.~\ref{fig_exp1}(c). Consequently, a tenfold bandwidth expansion ($B = $10/1 GHz = 10 GHz) can be achieved with a reconstruction error (NMSE) of 0.083. 

As mentioned in the previous section, the number of detectors $N$ affects the reconstruction performance. The NMSE decreased monotonically as a function of $N $ (Fig.~\ref{fig_exp1}(d)). However, increasing the filter's order attenuates high-frequency components further, burying them in noise and severely degrading the reconstruction. As shown in Fig.~\ref{fig_exp1}(e), fourth- or higher orders lead to significant high-frequency-component loss, resulting in a sharp increase in NMSE.

\begin{figure}[htbp]
\centering
\includegraphics[bb=0 0 624 408,width=14cm]{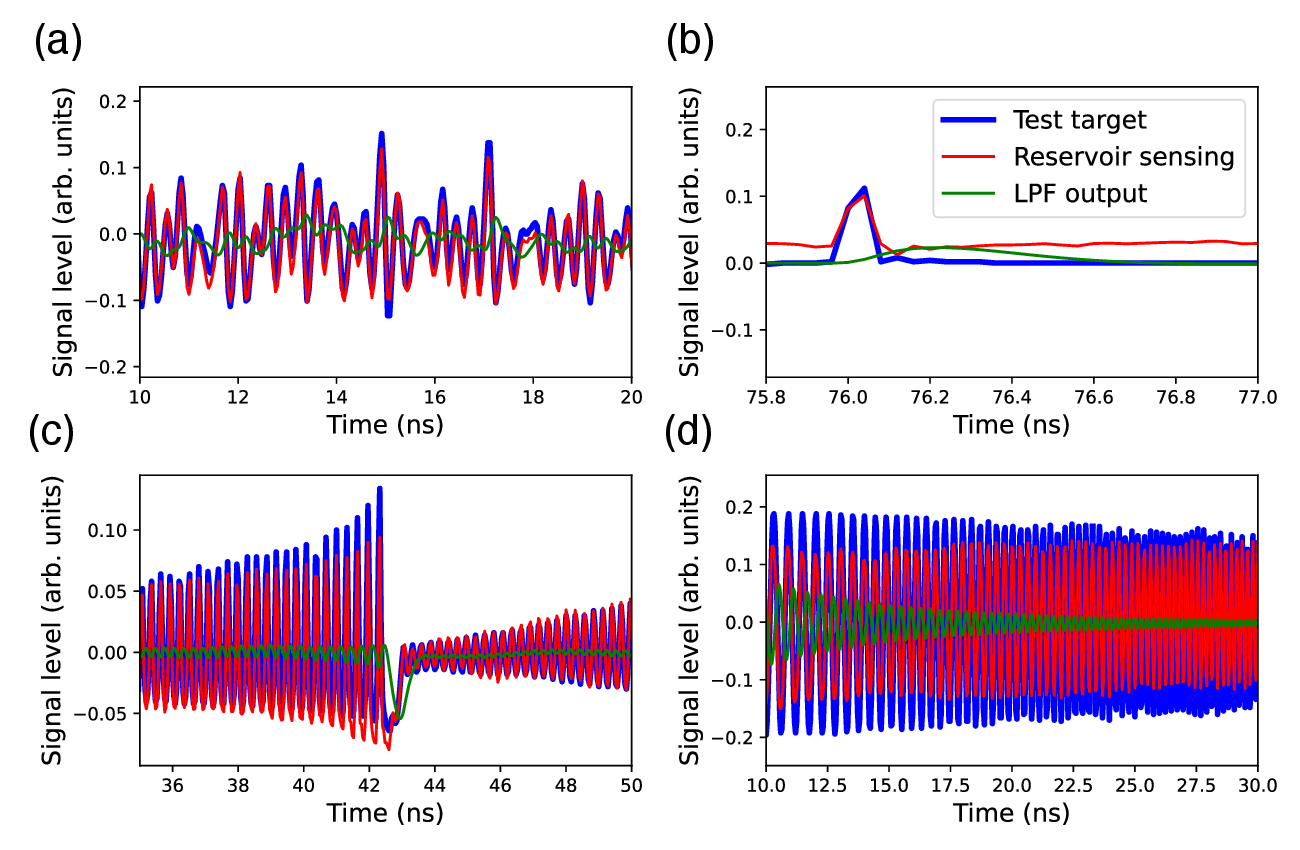}
\caption{Reconstruction results for (a) a laser chaos time-series, (b) a 100-ps pulse, (c) the Santa-Fe chaotic time-series, and (d) a chirped signal from 2 to 8 GHz.}
\label{fig_exp1-2}
\end{figure}

After training, the proposed approach can recover various waveforms, such as a laser chaos time-series (Fig.~\ref{fig_exp1-2}(a)), a 100-ps pulse (Fig.~\ref{fig_exp1-2}(b)), the Santa-Fe chaotic time-series \cite{Weigend1993}(Fig.~\ref{fig_exp1-2}(c)), and a chirped signal (Fig.~\ref{fig_exp1-2}(d)). See Supplementary Section~2 for detailed results. The reconstruction error can be further reduced by increasing the number of training samples. 

\subsection{Wavelength division multiplexing}
\begin{figure}[htbp]
\centering
\includegraphics[bb=0 0 570 490,width=14cm]{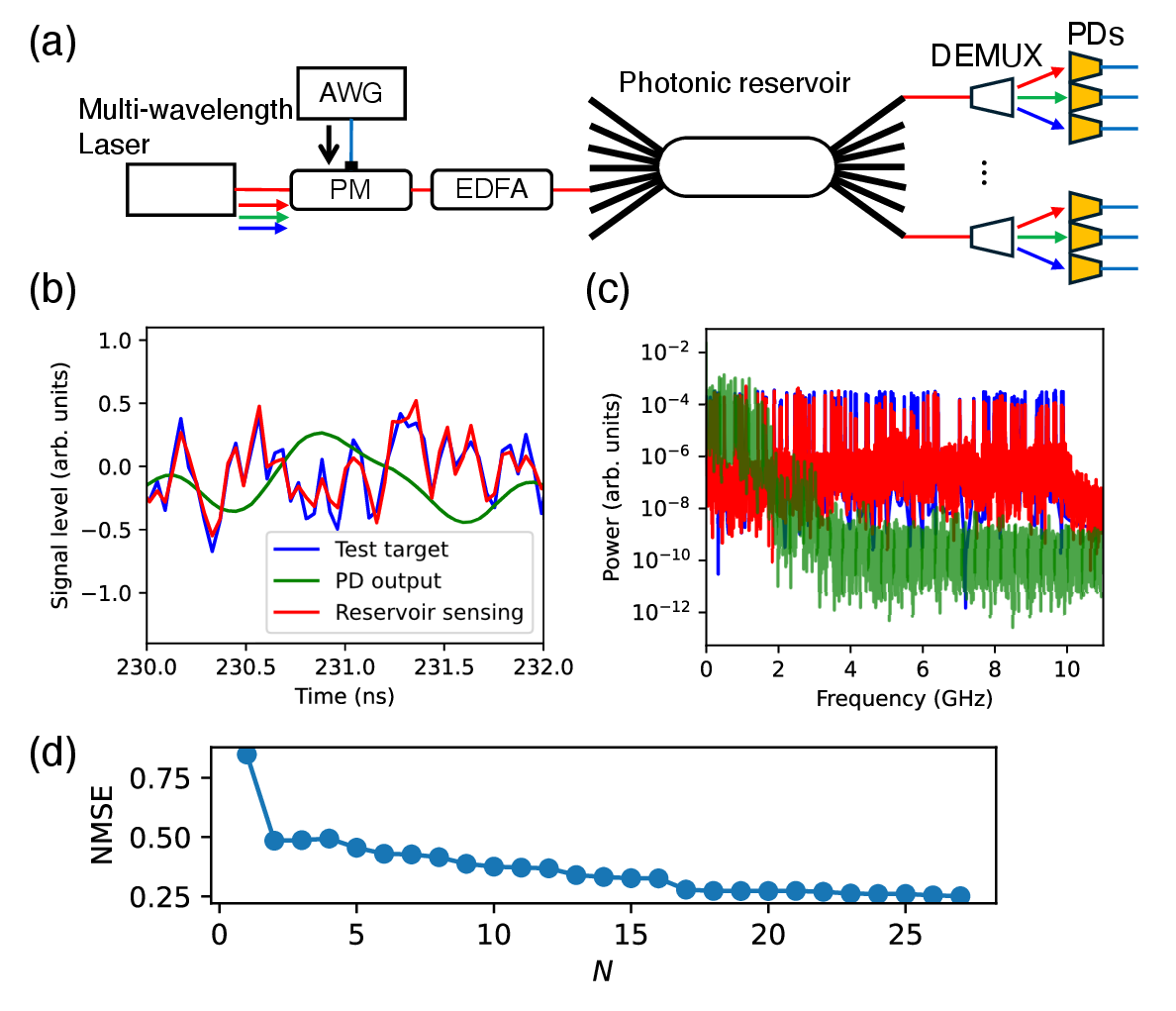}
\caption{WDM experimental setup and results. DEMUX: Demultiplexer, (arrayed waveguide grating).
(a) Conceptual schematic of the WDM-based sensing approach.
(b) Reconstructed 10-GHz waveform using narrowband photodetectors. 
(c) Power spectra of the original, detected, and reconstructed signals. 
(d) NMSE versus number of sensors $N$ obtained with WDM.}
\label{fig_wdm}
\end{figure}

One way of improving the reconstruction performance is to increase the number of reservoir nodes. However, in our current reservoir, the number of output ports is limited to 13. To overcome this limitation, we employed wavelength-division multiplexing (WDM). A photonic reservoir exhibits distinct responses at different input wavelengths \cite{Yamaguchi:2023aa}, as internal interference variations within the microcavity yield wavelength-dependent output features even for the same input signal. The WDM configuration is illustrated in Fig.~\ref{fig_wdm}.

For these experiments, we employed a narrowband photodetector (Newport/New Focus 1611; 3-dB bandwidth: 1.2 GHz). The goal was to demonstrate random waveform reconstruction with a 10-GHz bandwidth using the narrowband photodetectors. In the experiment shown in Fig.~\ref{fig_wdm}(a), we employed three wavelengths, namely, 1550.0, 1550.1, and 1550.2 nm, effectively corresponding to $N = 27$ equivalent channels, and obtained the reconstructed signal shown in Fig.~\ref{fig_wdm}(b). As shown in Fig.~\ref{fig_wdm}(c), the detected signals exhibit sharp frequency-component attenuation above 1.2 GHz, resembling the response of a fourth-order Butterworth filter.
Nevertheless, by employing WDM to increase $N$, we achieved a signal reconstruction with NMSE = 0.25. Figure~\ref{fig_wdm}(d) shows the NMSE as a function of $N$, where an NMSE below 0.30 can be observed for $N > 16$. These results reveal that the proposed approach is effective with practical photonic detection hardware, enabling signal recovery beyond the intrinsic detector bandwidth.

\section{Discussion and Conclusion}
In this paper, we propose and demonstrate a broadband signal detection framework using narrowband sensors, achieving more than an eightfold detectable bandwidth expansion. The main key of the proposed framework lies in the spatiotemporal encoding of broadband temporal information into narrowband reservoir channels, enabling high-frequency-component reconstruction, which is otherwise inaccessible to individual sensors. This approach allows narrowband detectors to collectively function as an effective broadband system and shows scalable improvements in reconstruction accuracy with increasing number of detectors, particularly at extremely high bandwidth ratios when sufficient training data are available. Further enhancements are possible via wavelength-division multiplexing.

Although our demonstration focused on 10-GHz signal recovery using $\sim$1-GHz detectors, the proposed approach is not restricted to this regime. The observed nearly tenfold bandwidth expansion suggests the possibility of detecting signals approaching 100 GHz with 10-GHz photodetectors, particularly when combined with high-speed photonic analog-to-digital converters \cite{Li:22,Yang:16,Xu_2016,photonics10070817,Godin31122022,10.1063/1.4941050,Mahjoubfar:2017aa,775471,Zhou:22,Yang:23}.

Beyond bandwidth expansion, the proposed detection scheme offers two unique features. (i) Owing to the intrinsic optical-phase sensitivity of the photonic reservoir, high-speed optical phase dynamics can be detected without complex interferometric setups, enabling simultaneous measurement of high-speed phase and intensity signals. (ii) The inherent nonlinearity of the photonic reservoir allows compensation for nonlinear distortions in photodetectors, leading to improved reconstruction compared with linear reservoirs (see See Supplementary Section~3 for details).

A drawback of the present implementation is the large optical leakage from the reservoir cavity, which results in significant scattering loss exceeding 15 dB (Fig.~\ref{fig_reservoir}). However, this strong scattering also enables multiple photodetectors to be placed around the cavity for enhanced detection capability, as shown in See Supplementary Section 4.

Our approach is promising for diverse applications, including optical quantum information processing, high-speed optical communications, and broadband spectroscopic sensing. Ultimately, the proposed framework paves the way for low-cost, energy-efficient broadband measurements, opening new horizons in ultrafast photonic technologies.

\begin{acknowledgements}
This work was supported in part by 
JSPS KAKENHI (Grant Nos.~JP22H05198, JP23K28157, JP22H05195, JP25H01129) and
JST CREST (Grant No.~JPMJCR24R2).
\end{acknowledgements}

%


\begin{thebibliography}{52}%
\makeatletter
\providecommand \@ifxundefined [1]{%
 \@ifx{#1\undefined}
}%
\providecommand \@ifnum [1]{%
 \ifnum #1\expandafter \@firstoftwo
 \else \expandafter \@secondoftwo
 \fi
}%
\providecommand \@ifx [1]{%
 \ifx #1\expandafter \@firstoftwo
 \else \expandafter \@secondoftwo
 \fi
}%
\providecommand \natexlab [1]{#1}%
\providecommand \enquote  [1]{``#1''}%
\providecommand \bibnamefont  [1]{#1}%
\providecommand \bibfnamefont [1]{#1}%
\providecommand \citenamefont [1]{#1}%
\providecommand \href@noop [0]{\@secondoftwo}%
\providecommand \href [0]{\begingroup \@sanitize@url \@href}%
\providecommand \@href[1]{\@@startlink{#1}\@@href}%
\providecommand \@@href[1]{\endgroup#1\@@endlink}%
\providecommand \@sanitize@url [0]{\catcode `\\12\catcode `\$12\catcode
  `\&12\catcode `\#12\catcode `\^12\catcode `\_12\catcode `\%12\relax}%
\providecommand \@@startlink[1]{}%
\providecommand \@@endlink[0]{}%
\providecommand \url  [0]{\begingroup\@sanitize@url \@url }%
\providecommand \@url [1]{\endgroup\@href {#1}{\urlprefix }}%
\providecommand \urlprefix  [0]{URL }%
\providecommand \Eprint [0]{\href }%
\providecommand \doibase [0]{https://doi.org/}%
\providecommand \selectlanguage [0]{\@gobble}%
\providecommand \bibinfo  [0]{\@secondoftwo}%
\providecommand \bibfield  [0]{\@secondoftwo}%
\providecommand \translation [1]{[#1]}%
\providecommand \BibitemOpen [0]{}%
\providecommand \bibitemStop [0]{}%
\providecommand \bibitemNoStop [0]{.\EOS\space}%
\providecommand \EOS [0]{\spacefactor3000\relax}%
\providecommand \BibitemShut  [1]{\csname bibitem#1\endcsname}%
\let\auto@bib@innerbib\@empty
\bibitem [{\citenamefont {Mueller}\ \emph {et~al.}(2010)\citenamefont
  {Mueller}, \citenamefont {Xia},\ and\ \citenamefont
  {Avouris}}]{Mueller:2010aa}%
  \BibitemOpen
  \bibfield  {author} {\bibinfo {author} {\bibfnamefont {T.}~\bibnamefont
  {Mueller}}, \bibinfo {author} {\bibfnamefont {F.}~\bibnamefont {Xia}},\ and\
  \bibinfo {author} {\bibfnamefont {P.}~\bibnamefont {Avouris}},\ }\bibfield
  {title} {\bibinfo {title} {Graphene photodetectors for high-speed optical
  communications},\ }\href {https://doi.org/10.1038/nphoton.2010.40} {\bibfield
   {journal} {\bibinfo  {journal} {Nature Photonics}\ }\textbf {\bibinfo
  {volume} {4}},\ \bibinfo {pages} {297} (\bibinfo {year} {2010})}\BibitemShut
  {NoStop}%
\bibitem [{\citenamefont {Liu}\ \emph {et~al.}(2025)\citenamefont {Liu},
  \citenamefont {Yang}, \citenamefont {Li},\ and\ \citenamefont
  {Tong}}]{Liu:25}%
  \BibitemOpen
  \bibfield  {author} {\bibinfo {author} {\bibfnamefont {T.}~\bibnamefont
  {Liu}}, \bibinfo {author} {\bibfnamefont {G.}~\bibnamefont {Yang}}, \bibinfo
  {author} {\bibfnamefont {J.}~\bibnamefont {Li}},\ and\ \bibinfo {author}
  {\bibfnamefont {C.}~\bibnamefont {Tong}},\ }\bibfield  {title} {\bibinfo
  {title} {High-speed avalanche photodiodes for optical communication},\ }\href
  {https://doi.org/10.1364/PRJ.544561} {\bibfield  {journal} {\bibinfo
  {journal} {Photon. Res.}\ }\textbf {\bibinfo {volume} {13}},\ \bibinfo
  {pages} {1438} (\bibinfo {year} {2025})}\BibitemShut {NoStop}%
\bibitem [{\citenamefont {Zhao}\ \emph {et~al.}(2017)\citenamefont {Zhao},
  \citenamefont {Liu}, \citenamefont {Liu},\ and\ \citenamefont
  {Zhu}}]{Zhao_2017}%
  \BibitemOpen
  \bibfield  {author} {\bibinfo {author} {\bibfnamefont {Z.}~\bibnamefont
  {Zhao}}, \bibinfo {author} {\bibfnamefont {J.}~\bibnamefont {Liu}}, \bibinfo
  {author} {\bibfnamefont {Y.}~\bibnamefont {Liu}},\ and\ \bibinfo {author}
  {\bibfnamefont {N.}~\bibnamefont {Zhu}},\ }\bibfield  {title} {\bibinfo
  {title} {High-speed photodetectors in optical communication system *},\
  }\href {https://doi.org/10.1088/1674-4926/38/12/121001} {\bibfield  {journal}
  {\bibinfo  {journal} {Journal of Semiconductors}\ }\textbf {\bibinfo {volume}
  {38}},\ \bibinfo {pages} {121001} (\bibinfo {year} {2017})}\BibitemShut
  {NoStop}%
\bibitem [{\citenamefont {Chen}\ \emph {et~al.}(2021)\citenamefont {Chen},
  \citenamefont {Chen},\ and\ \citenamefont {Deng}}]{photonics8010014}%
  \BibitemOpen
  \bibfield  {author} {\bibinfo {author} {\bibfnamefont {B.}~\bibnamefont
  {Chen}}, \bibinfo {author} {\bibfnamefont {Y.}~\bibnamefont {Chen}},\ and\
  \bibinfo {author} {\bibfnamefont {Z.}~\bibnamefont {Deng}},\ }\bibfield
  {title} {\bibinfo {title} {Recent advances in high speed photodetectors for
  eswir/mwir/lwir applications},\ }\bibfield  {journal} {\bibinfo  {journal}
  {Photonics}\ }\textbf {\bibinfo {volume} {8}},\ \href
  {https://doi.org/10.3390/photonics8010014} {10.3390/photonics8010014}
  (\bibinfo {year} {2021})\BibitemShut {NoStop}%
\bibitem [{\citenamefont {Goda}\ \emph {et~al.}(2009)\citenamefont {Goda},
  \citenamefont {Tsia},\ and\ \citenamefont {Jalali}}]{Goda:2009aa}%
  \BibitemOpen
  \bibfield  {author} {\bibinfo {author} {\bibfnamefont {K.}~\bibnamefont
  {Goda}}, \bibinfo {author} {\bibfnamefont {K.~K.}\ \bibnamefont {Tsia}},\
  and\ \bibinfo {author} {\bibfnamefont {B.}~\bibnamefont {Jalali}},\
  }\bibfield  {title} {\bibinfo {title} {Serial time-encoded amplified imaging
  for real-time observation of fast dynamic phenomena},\ }\href
  {https://doi.org/10.1038/nature07980} {\bibfield  {journal} {\bibinfo
  {journal} {Nature}\ }\textbf {\bibinfo {volume} {458}},\ \bibinfo {pages}
  {1145} (\bibinfo {year} {2009})}\BibitemShut {NoStop}%
\bibitem [{\citenamefont {Gibson}\ \emph {et~al.}(2020)\citenamefont {Gibson},
  \citenamefont {Johnson},\ and\ \citenamefont {Padgett}}]{Gibson:20}%
  \BibitemOpen
  \bibfield  {author} {\bibinfo {author} {\bibfnamefont {G.~M.}\ \bibnamefont
  {Gibson}}, \bibinfo {author} {\bibfnamefont {S.~D.}\ \bibnamefont
  {Johnson}},\ and\ \bibinfo {author} {\bibfnamefont {M.~J.}\ \bibnamefont
  {Padgett}},\ }\bibfield  {title} {\bibinfo {title} {Single-pixel imaging 12
  years on: a review},\ }\href {https://doi.org/10.1364/OE.403195} {\bibfield
  {journal} {\bibinfo  {journal} {Opt. Express}\ }\textbf {\bibinfo {volume}
  {28}},\ \bibinfo {pages} {28190} (\bibinfo {year} {2020})}\BibitemShut
  {NoStop}%
\bibitem [{\citenamefont {Chen}\ \emph {et~al.}(2024)\citenamefont {Chen},
  \citenamefont {Zhang}, \citenamefont {Sharawi},\ and\ \citenamefont
  {Kashyap}}]{app14083410}%
  \BibitemOpen
  \bibfield  {author} {\bibinfo {author} {\bibfnamefont {Q.}~\bibnamefont
  {Chen}}, \bibinfo {author} {\bibfnamefont {X.}~\bibnamefont {Zhang}},
  \bibinfo {author} {\bibfnamefont {M.~S.}\ \bibnamefont {Sharawi}},\ and\
  \bibinfo {author} {\bibfnamefont {R.}~\bibnamefont {Kashyap}},\ }\bibfield
  {title} {\bibinfo {title} {Advances in high--speed, high--power photodiodes:
  From fundamentals to applications},\ }\bibfield  {journal} {\bibinfo
  {journal} {Applied Sciences}\ }\textbf {\bibinfo {volume} {14}},\ \href
  {https://doi.org/10.3390/app14083410} {10.3390/app14083410} (\bibinfo {year}
  {2024})\BibitemShut {NoStop}%
\bibitem [{\citenamefont {Koch}\ \emph {et~al.}(2023)\citenamefont {Koch},
  \citenamefont {Mittleman}, \citenamefont {Ornik},\ and\ \citenamefont
  {Castro-Camus}}]{Koch:2023aa}%
  \BibitemOpen
  \bibfield  {author} {\bibinfo {author} {\bibfnamefont {M.}~\bibnamefont
  {Koch}}, \bibinfo {author} {\bibfnamefont {D.~M.}\ \bibnamefont {Mittleman}},
  \bibinfo {author} {\bibfnamefont {J.}~\bibnamefont {Ornik}},\ and\ \bibinfo
  {author} {\bibfnamefont {E.}~\bibnamefont {Castro-Camus}},\ }\bibfield
  {title} {\bibinfo {title} {Terahertz time-domain spectroscopy},\ }\href
  {https://doi.org/10.1038/s43586-023-00232-z} {\bibfield  {journal} {\bibinfo
  {journal} {Nature Reviews Methods Primers}\ }\textbf {\bibinfo {volume}
  {3}},\ \bibinfo {pages} {48} (\bibinfo {year} {2023})}\BibitemShut {NoStop}%
\bibitem [{\citenamefont {Chen}\ \emph {et~al.}(2023)\citenamefont {Chen},
  \citenamefont {Sludds}, \citenamefont {Davis}, \citenamefont {Christen},
  \citenamefont {Bernstein}, \citenamefont {Ateshian}, \citenamefont {Heuser},
  \citenamefont {Heermeier}, \citenamefont {Lott}, \citenamefont
  {Reitzenstein}, \citenamefont {Hamerly},\ and\ \citenamefont
  {Englund}}]{Chen:2023wi}%
  \BibitemOpen
  \bibfield  {author} {\bibinfo {author} {\bibfnamefont {Z.}~\bibnamefont
  {Chen}}, \bibinfo {author} {\bibfnamefont {A.}~\bibnamefont {Sludds}},
  \bibinfo {author} {\bibfnamefont {R.}~\bibnamefont {Davis}}, \bibinfo
  {author} {\bibfnamefont {I.}~\bibnamefont {Christen}}, \bibinfo {author}
  {\bibfnamefont {L.}~\bibnamefont {Bernstein}}, \bibinfo {author}
  {\bibfnamefont {L.}~\bibnamefont {Ateshian}}, \bibinfo {author}
  {\bibfnamefont {T.}~\bibnamefont {Heuser}}, \bibinfo {author} {\bibfnamefont
  {N.}~\bibnamefont {Heermeier}}, \bibinfo {author} {\bibfnamefont {J.~A.}\
  \bibnamefont {Lott}}, \bibinfo {author} {\bibfnamefont {S.}~\bibnamefont
  {Reitzenstein}}, \bibinfo {author} {\bibfnamefont {R.}~\bibnamefont
  {Hamerly}},\ and\ \bibinfo {author} {\bibfnamefont {D.}~\bibnamefont
  {Englund}},\ }\bibfield  {title} {\bibinfo {title} {Deep learning with
  coherent vcsel neural networks},\ }\href
  {https://doi.org/10.1038/s41566-023-01233-w} {\bibfield  {journal} {\bibinfo
  {journal} {Nature Photonics}\ }\textbf {\bibinfo {volume} {17}},\ \bibinfo
  {pages} {723} (\bibinfo {year} {2023})}\BibitemShut {NoStop}%
\bibitem [{\citenamefont {Freire}\ \emph {et~al.}(2023)\citenamefont {Freire},
  \citenamefont {Manuylovich}, \citenamefont {Prilepsky},\ and\ \citenamefont
  {Turitsyn}}]{Freire:23}%
  \BibitemOpen
  \bibfield  {author} {\bibinfo {author} {\bibfnamefont {P.}~\bibnamefont
  {Freire}}, \bibinfo {author} {\bibfnamefont {E.}~\bibnamefont {Manuylovich}},
  \bibinfo {author} {\bibfnamefont {J.~E.}\ \bibnamefont {Prilepsky}},\ and\
  \bibinfo {author} {\bibfnamefont {S.~K.}\ \bibnamefont {Turitsyn}},\
  }\bibfield  {title} {\bibinfo {title} {Artificial neural networks for
  photonic applications---from algorithms to implementation: tutorial},\ }\href
  {https://doi.org/10.1364/AOP.484119} {\bibfield  {journal} {\bibinfo
  {journal} {Adv. Opt. Photon.}\ }\textbf {\bibinfo {volume} {15}},\ \bibinfo
  {pages} {739} (\bibinfo {year} {2023})}\BibitemShut {NoStop}%
\bibitem [{\citenamefont {Kawasaki}\ \emph {et~al.}(2024)\citenamefont
  {Kawasaki}, \citenamefont {Ide}, \citenamefont {Brunel}, \citenamefont
  {Suzuki}, \citenamefont {Nehra}, \citenamefont {Nakashima}, \citenamefont
  {Kashiwazaki}, \citenamefont {Inoue}, \citenamefont {Umeki}, \citenamefont
  {China}, \citenamefont {Yabuno}, \citenamefont {Miki}, \citenamefont {Terai},
  \citenamefont {Yamashima}, \citenamefont {Sakaguchi}, \citenamefont {Takase},
  \citenamefont {Endo}, \citenamefont {Asavanant},\ and\ \citenamefont
  {Furusawa}}]{Kawasaki:2024aa}%
  \BibitemOpen
  \bibfield  {author} {\bibinfo {author} {\bibfnamefont {A.}~\bibnamefont
  {Kawasaki}}, \bibinfo {author} {\bibfnamefont {R.}~\bibnamefont {Ide}},
  \bibinfo {author} {\bibfnamefont {H.}~\bibnamefont {Brunel}}, \bibinfo
  {author} {\bibfnamefont {T.}~\bibnamefont {Suzuki}}, \bibinfo {author}
  {\bibfnamefont {R.}~\bibnamefont {Nehra}}, \bibinfo {author} {\bibfnamefont
  {K.}~\bibnamefont {Nakashima}}, \bibinfo {author} {\bibfnamefont
  {T.}~\bibnamefont {Kashiwazaki}}, \bibinfo {author} {\bibfnamefont
  {A.}~\bibnamefont {Inoue}}, \bibinfo {author} {\bibfnamefont
  {T.}~\bibnamefont {Umeki}}, \bibinfo {author} {\bibfnamefont
  {F.}~\bibnamefont {China}}, \bibinfo {author} {\bibfnamefont
  {M.}~\bibnamefont {Yabuno}}, \bibinfo {author} {\bibfnamefont
  {S.}~\bibnamefont {Miki}}, \bibinfo {author} {\bibfnamefont {H.}~\bibnamefont
  {Terai}}, \bibinfo {author} {\bibfnamefont {T.}~\bibnamefont {Yamashima}},
  \bibinfo {author} {\bibfnamefont {A.}~\bibnamefont {Sakaguchi}}, \bibinfo
  {author} {\bibfnamefont {K.}~\bibnamefont {Takase}}, \bibinfo {author}
  {\bibfnamefont {M.}~\bibnamefont {Endo}}, \bibinfo {author} {\bibfnamefont
  {W.}~\bibnamefont {Asavanant}},\ and\ \bibinfo {author} {\bibfnamefont
  {A.}~\bibnamefont {Furusawa}},\ }\bibfield  {title} {\bibinfo {title}
  {Broadband generation and tomography of non-gaussian states for ultra-fast
  optical quantum processors},\ }\href
  {https://doi.org/10.1038/s41467-024-53408-w} {\bibfield  {journal} {\bibinfo
  {journal} {Nature Communications}\ }\textbf {\bibinfo {volume} {15}},\
  \bibinfo {pages} {9075} (\bibinfo {year} {2024})}\BibitemShut {NoStop}%
\bibitem [{\citenamefont {Marpaung}\ \emph {et~al.}(2019)\citenamefont
  {Marpaung}, \citenamefont {Yao},\ and\ \citenamefont
  {Capmany}}]{Marpaung:2019aa}%
  \BibitemOpen
  \bibfield  {author} {\bibinfo {author} {\bibfnamefont {D.}~\bibnamefont
  {Marpaung}}, \bibinfo {author} {\bibfnamefont {J.}~\bibnamefont {Yao}},\ and\
  \bibinfo {author} {\bibfnamefont {J.}~\bibnamefont {Capmany}},\ }\bibfield
  {title} {\bibinfo {title} {Integrated microwave photonics},\ }\href
  {https://doi.org/10.1038/s41566-018-0310-5} {\bibfield  {journal} {\bibinfo
  {journal} {Nature Photonics}\ }\textbf {\bibinfo {volume} {13}},\ \bibinfo
  {pages} {80} (\bibinfo {year} {2019})}\BibitemShut {NoStop}%
\bibitem [{\citenamefont {Rajabali}\ and\ \citenamefont
  {Benea-Chelmus}()}]{10.1063/5.0146912}%
  \BibitemOpen
  \bibfield  {author} {\bibinfo {author} {\bibfnamefont {S.}~\bibnamefont
  {Rajabali}}\ and\ \bibinfo {author} {\bibfnamefont {I.-C.}\ \bibnamefont
  {Benea-Chelmus}}\ }\href {https://doi.org/10.1063/5.0146912}
  {10.1063/5.0146912}\BibitemShut {NoStop}%
\bibitem [{\citenamefont {Han}\ \emph {et~al.}(2025)\citenamefont {Han},
  \citenamefont {Tuo}, \citenamefont {Deng}, \citenamefont {Han}, \citenamefont
  {He}, \citenamefont {Han}, \citenamefont {Guo}, \citenamefont {Zhou},
  \citenamefont {Yu}, \citenamefont {Gou}, \citenamefont {Li}, \citenamefont
  {Lu},\ and\ \citenamefont {Wang}}]{Han:2025_LPR}%
  \BibitemOpen
  \bibfield  {author} {\bibinfo {author} {\bibfnamefont {J.}~\bibnamefont
  {Han}}, \bibinfo {author} {\bibfnamefont {T.}~\bibnamefont {Tuo}}, \bibinfo
  {author} {\bibfnamefont {W.}~\bibnamefont {Deng}}, \bibinfo {author}
  {\bibfnamefont {X.}~\bibnamefont {Han}}, \bibinfo {author} {\bibfnamefont
  {M.}~\bibnamefont {He}}, \bibinfo {author} {\bibfnamefont {C.}~\bibnamefont
  {Han}}, \bibinfo {author} {\bibfnamefont {L.}~\bibnamefont {Guo}}, \bibinfo
  {author} {\bibfnamefont {H.}~\bibnamefont {Zhou}}, \bibinfo {author}
  {\bibfnamefont {H.}~\bibnamefont {Yu}}, \bibinfo {author} {\bibfnamefont
  {J.}~\bibnamefont {Gou}}, \bibinfo {author} {\bibfnamefont {G.}~\bibnamefont
  {Li}}, \bibinfo {author} {\bibfnamefont {D.}~\bibnamefont {Lu}},\ and\
  \bibinfo {author} {\bibfnamefont {J.}~\bibnamefont {Wang}},\ }\bibfield
  {title} {\bibinfo {title} {2d/organic photovoltage field-effect
  transistors},\ }\href
  {https://doi.org/https://doi.org/10.1002/lpor.202500268} {\bibfield
  {journal} {\bibinfo  {journal} {Laser \& Photonics Reviews}\ }\textbf
  {\bibinfo {volume} {19}},\ \bibinfo {pages} {2500268} (\bibinfo {year}
  {2025})},\ \Eprint
  {https://arxiv.org/abs/https://onlinelibrary.wiley.com/doi/pdf/10.1002/lpor.202500268}
  {https://onlinelibrary.wiley.com/doi/pdf/10.1002/lpor.202500268} \BibitemShut
  {NoStop}%
\bibitem [{\citenamefont {Sorger}\ and\ \citenamefont
  {Maiti}(2020)}]{Sorger:20}%
  \BibitemOpen
  \bibfield  {author} {\bibinfo {author} {\bibfnamefont {V.~J.}\ \bibnamefont
  {Sorger}}\ and\ \bibinfo {author} {\bibfnamefont {R.}~\bibnamefont {Maiti}},\
  }\bibfield  {title} {\bibinfo {title} {Roadmap for gain-bandwidth-product
  enhanced photodetectors: opinion},\ }\href
  {https://doi.org/10.1364/OME.400423} {\bibfield  {journal} {\bibinfo
  {journal} {Opt. Mater. Express}\ }\textbf {\bibinfo {volume} {10}},\ \bibinfo
  {pages} {2192} (\bibinfo {year} {2020})}\BibitemShut {NoStop}%
\bibitem [{\citenamefont {Zhu}\ \emph {et~al.}(2025)\citenamefont {Zhu},
  \citenamefont {Wu}, \citenamefont {Zhang},\ and\ \citenamefont
  {Xu}}]{Zhu:25}%
  \BibitemOpen
  \bibfield  {author} {\bibinfo {author} {\bibfnamefont {S.}~\bibnamefont
  {Zhu}}, \bibinfo {author} {\bibfnamefont {B.}~\bibnamefont {Wu}}, \bibinfo
  {author} {\bibfnamefont {Y.}~\bibnamefont {Zhang}},\ and\ \bibinfo {author}
  {\bibfnamefont {Y.}~\bibnamefont {Xu}},\ }\bibfield  {title} {\bibinfo
  {title} {Design and simulation of a high gain--bandwidth product
  ingaas/algaassb avalanche photodiode with a p-type hybrid absorption layer},\
  }\href {https://doi.org/10.1364/AO.551928} {\bibfield  {journal} {\bibinfo
  {journal} {Appl. Opt.}\ }\textbf {\bibinfo {volume} {64}},\ \bibinfo {pages}
  {3229} (\bibinfo {year} {2025})}\BibitemShut {NoStop}%
\bibitem [{\citenamefont {Li}\ \emph {et~al.}(2022{\natexlab{a}})\citenamefont
  {Li}, \citenamefont {Chen}, \citenamefont {Zhao}, \citenamefont {Liu},\ and\
  \citenamefont {Zhao}}]{doi:10.1126/sciadv.abq0187}%
  \BibitemOpen
  \bibfield  {author} {\bibinfo {author} {\bibfnamefont {Y.}~\bibnamefont
  {Li}}, \bibinfo {author} {\bibfnamefont {G.}~\bibnamefont {Chen}}, \bibinfo
  {author} {\bibfnamefont {S.}~\bibnamefont {Zhao}}, \bibinfo {author}
  {\bibfnamefont {C.}~\bibnamefont {Liu}},\ and\ \bibinfo {author}
  {\bibfnamefont {N.}~\bibnamefont {Zhao}},\ }\bibfield  {title} {\bibinfo
  {title} {Addressing gain-bandwidth trade-off by a monolithically integrated
  photovoltaic transistor},\ }\href {https://doi.org/10.1126/sciadv.abq0187}
  {\bibfield  {journal} {\bibinfo  {journal} {Science Advances}\ }\textbf
  {\bibinfo {volume} {8}},\ \bibinfo {pages} {eabq0187} (\bibinfo {year}
  {2022}{\natexlab{a}})},\ \Eprint
  {https://arxiv.org/abs/https://www.science.org/doi/pdf/10.1126/sciadv.abq0187}
  {https://www.science.org/doi/pdf/10.1126/sciadv.abq0187} \BibitemShut
  {NoStop}%
\bibitem [{\citenamefont {Zhou}\ and\ \citenamefont {Chee}(2019)}]{Chee19}%
  \BibitemOpen
  \bibfield  {author} {\bibinfo {author} {\bibfnamefont {T.}~\bibnamefont
  {Zhou}}\ and\ \bibinfo {author} {\bibfnamefont {K.~W.}\ \bibnamefont
  {Chee}},\ }\bibfield  {title} {\bibinfo {title} {Overcoming the
  bandwidth-quantum efficiency trade-off in conventional photodetectors},\ }in\
  \href {https://doi.org/10.5772/intechopen.86506} {\emph {\bibinfo {booktitle}
  {Advances in Photodetectors - Research and Applications}}},\ \bibinfo
  {editor} {edited by\ \bibinfo {editor} {\bibfnamefont {K.}~\bibnamefont
  {Chee}}}\ (\bibinfo  {publisher} {IntechOpen},\ \bibinfo {address} {London},\
  \bibinfo {year} {2019})\ Chap.~\bibinfo {chapter} {7}\BibitemShut {NoStop}%
\bibitem [{\citenamefont {Song}\ \emph {et~al.}(2014)\citenamefont {Song},
  \citenamefont {Eu-Jin}, \citenamefont {Luo}, \citenamefont {Huang},
  \citenamefont {Tu}, \citenamefont {Jia}, \citenamefont {Fang}, \citenamefont
  {Liow}, \citenamefont {Yu},\ and\ \citenamefont {Lo}}]{Song:14}%
  \BibitemOpen
  \bibfield  {author} {\bibinfo {author} {\bibfnamefont {J.}~\bibnamefont
  {Song}}, \bibinfo {author} {\bibfnamefont {A.~L.}\ \bibnamefont {Eu-Jin}},
  \bibinfo {author} {\bibfnamefont {X.}~\bibnamefont {Luo}}, \bibinfo {author}
  {\bibfnamefont {Y.}~\bibnamefont {Huang}}, \bibinfo {author} {\bibfnamefont
  {X.}~\bibnamefont {Tu}}, \bibinfo {author} {\bibfnamefont {L.}~\bibnamefont
  {Jia}}, \bibinfo {author} {\bibfnamefont {Q.}~\bibnamefont {Fang}}, \bibinfo
  {author} {\bibfnamefont {T.-Y.}\ \bibnamefont {Liow}}, \bibinfo {author}
  {\bibfnamefont {M.}~\bibnamefont {Yu}},\ and\ \bibinfo {author}
  {\bibfnamefont {G.-Q.}\ \bibnamefont {Lo}},\ }\bibfield  {title} {\bibinfo
  {title} {A microring resonator photodetector for enhancement in l-band
  performance},\ }\href {https://doi.org/10.1364/OE.22.026976} {\bibfield
  {journal} {\bibinfo  {journal} {Opt. Express}\ }\textbf {\bibinfo {volume}
  {22}},\ \bibinfo {pages} {26976} (\bibinfo {year} {2014})}\BibitemShut
  {NoStop}%
\bibitem [{\citenamefont {Kato}(1999)}]{775466}%
  \BibitemOpen
  \bibfield  {author} {\bibinfo {author} {\bibfnamefont {K.}~\bibnamefont
  {Kato}},\ }\bibfield  {title} {\bibinfo {title}
  {Ultrawide-band/high-frequency photodetectors},\ }\href
  {https://doi.org/10.1109/22.775466} {\bibfield  {journal} {\bibinfo
  {journal} {IEEE Transactions on Microwave Theory and Techniques}\ }\textbf
  {\bibinfo {volume} {47}},\ \bibinfo {pages} {1265} (\bibinfo {year}
  {1999})}\BibitemShut {NoStop}%
\bibitem [{\citenamefont {Giboney}\ \emph {et~al.}(1992)\citenamefont
  {Giboney}, \citenamefont {Rodwell},\ and\ \citenamefont {Bowers}}]{180577}%
  \BibitemOpen
  \bibfield  {author} {\bibinfo {author} {\bibfnamefont {K.}~\bibnamefont
  {Giboney}}, \bibinfo {author} {\bibfnamefont {M.}~\bibnamefont {Rodwell}},\
  and\ \bibinfo {author} {\bibfnamefont {J.}~\bibnamefont {Bowers}},\
  }\bibfield  {title} {\bibinfo {title} {Traveling-wave photodetectors},\
  }\href {https://doi.org/10.1109/68.180577} {\bibfield  {journal} {\bibinfo
  {journal} {IEEE Photonics Technology Letters}\ }\textbf {\bibinfo {volume}
  {4}},\ \bibinfo {pages} {1363} (\bibinfo {year} {1992})}\BibitemShut
  {NoStop}%
\bibitem [{\citenamefont {Bowers}\ and\ \citenamefont
  {Burrus}(1987)}]{1075419}%
  \BibitemOpen
  \bibfield  {author} {\bibinfo {author} {\bibfnamefont {J.}~\bibnamefont
  {Bowers}}\ and\ \bibinfo {author} {\bibfnamefont {C.}~\bibnamefont
  {Burrus}},\ }\bibfield  {title} {\bibinfo {title} {Ultrawide-band
  long-wavelength p-i-n photodetectors},\ }\href
  {https://doi.org/10.1109/JLT.1987.1075419} {\bibfield  {journal} {\bibinfo
  {journal} {Journal of Lightwave Technology}\ }\textbf {\bibinfo {volume}
  {5}},\ \bibinfo {pages} {1339} (\bibinfo {year} {1987})}\BibitemShut
  {NoStop}%
\bibitem [{\citenamefont {Gao}\ \emph {et~al.}(2017)\citenamefont {Gao},
  \citenamefont {Cansizoglu}, \citenamefont {Polat}, \citenamefont
  {Ghandiparsi}, \citenamefont {Kaya}, \citenamefont {Mamtaz}, \citenamefont
  {Mayet}, \citenamefont {Wang}, \citenamefont {Zhang}, \citenamefont {Yamada},
  \citenamefont {Devine}, \citenamefont {Elrefaie}, \citenamefont {Wang},\ and\
  \citenamefont {Islam}}]{Gao:2017aa}%
  \BibitemOpen
  \bibfield  {author} {\bibinfo {author} {\bibfnamefont {Y.}~\bibnamefont
  {Gao}}, \bibinfo {author} {\bibfnamefont {H.}~\bibnamefont {Cansizoglu}},
  \bibinfo {author} {\bibfnamefont {K.~G.}\ \bibnamefont {Polat}}, \bibinfo
  {author} {\bibfnamefont {S.}~\bibnamefont {Ghandiparsi}}, \bibinfo {author}
  {\bibfnamefont {A.}~\bibnamefont {Kaya}}, \bibinfo {author} {\bibfnamefont
  {H.~H.}\ \bibnamefont {Mamtaz}}, \bibinfo {author} {\bibfnamefont {A.~S.}\
  \bibnamefont {Mayet}}, \bibinfo {author} {\bibfnamefont {Y.}~\bibnamefont
  {Wang}}, \bibinfo {author} {\bibfnamefont {X.}~\bibnamefont {Zhang}},
  \bibinfo {author} {\bibfnamefont {T.}~\bibnamefont {Yamada}}, \bibinfo
  {author} {\bibfnamefont {E.~P.}\ \bibnamefont {Devine}}, \bibinfo {author}
  {\bibfnamefont {A.~F.}\ \bibnamefont {Elrefaie}}, \bibinfo {author}
  {\bibfnamefont {S.-Y.}\ \bibnamefont {Wang}},\ and\ \bibinfo {author}
  {\bibfnamefont {M.~S.}\ \bibnamefont {Islam}},\ }\bibfield  {title} {\bibinfo
  {title} {Photon-trapping microstructures enable high-speed high-efficiency
  silicon photodiodes},\ }\href {https://doi.org/10.1038/nphoton.2017.37}
  {\bibfield  {journal} {\bibinfo  {journal} {Nature Photonics}\ }\textbf
  {\bibinfo {volume} {11}},\ \bibinfo {pages} {301} (\bibinfo {year}
  {2017})}\BibitemShut {NoStop}%
\bibitem [{\citenamefont {Li}\ \emph {et~al.}(2022{\natexlab{b}})\citenamefont
  {Li}, \citenamefont {Wang}, \citenamefont {Zhang}, \citenamefont {Shang},
  \citenamefont {Lyu}, \citenamefont {Lyu}, \citenamefont {Zeng}, \citenamefont
  {Zhang}, \citenamefont {Zhang}, \citenamefont {Li}, \citenamefont {Xia},\
  and\ \citenamefont {Liu}}]{Li:22}%
  \BibitemOpen
  \bibfield  {author} {\bibinfo {author} {\bibfnamefont {Z.}~\bibnamefont
  {Li}}, \bibinfo {author} {\bibfnamefont {X.}~\bibnamefont {Wang}}, \bibinfo
  {author} {\bibfnamefont {Y.}~\bibnamefont {Zhang}}, \bibinfo {author}
  {\bibfnamefont {C.}~\bibnamefont {Shang}}, \bibinfo {author} {\bibfnamefont
  {W.}~\bibnamefont {Lyu}}, \bibinfo {author} {\bibfnamefont {Y.}~\bibnamefont
  {Lyu}}, \bibinfo {author} {\bibfnamefont {C.}~\bibnamefont {Zeng}}, \bibinfo
  {author} {\bibfnamefont {Z.}~\bibnamefont {Zhang}}, \bibinfo {author}
  {\bibfnamefont {S.}~\bibnamefont {Zhang}}, \bibinfo {author} {\bibfnamefont
  {H.}~\bibnamefont {Li}}, \bibinfo {author} {\bibfnamefont {J.}~\bibnamefont
  {Xia}},\ and\ \bibinfo {author} {\bibfnamefont {Y.}~\bibnamefont {Liu}},\
  }\bibfield  {title} {\bibinfo {title} {Photonic sampling analog-to-digital
  conversion based on time and wavelength interleaved ultra-short optical pulse
  train generated by using monolithic integrated lnoi intensity and phase
  modulator},\ }\href {https://doi.org/10.1364/OE.465733} {\bibfield  {journal}
  {\bibinfo  {journal} {Opt. Express}\ }\textbf {\bibinfo {volume} {30}},\
  \bibinfo {pages} {29611} (\bibinfo {year} {2022}{\natexlab{b}})}\BibitemShut
  {NoStop}%
\bibitem [{\citenamefont {Yang}\ \emph {et~al.}(2016)\citenamefont {Yang},
  \citenamefont {Zou}, \citenamefont {Yu}, \citenamefont {Wu},\ and\
  \citenamefont {Chen}}]{Yang:16}%
  \BibitemOpen
  \bibfield  {author} {\bibinfo {author} {\bibfnamefont {G.}~\bibnamefont
  {Yang}}, \bibinfo {author} {\bibfnamefont {W.}~\bibnamefont {Zou}}, \bibinfo
  {author} {\bibfnamefont {L.}~\bibnamefont {Yu}}, \bibinfo {author}
  {\bibfnamefont {K.}~\bibnamefont {Wu}},\ and\ \bibinfo {author}
  {\bibfnamefont {J.}~\bibnamefont {Chen}},\ }\bibfield  {title} {\bibinfo
  {title} {Compensation of multi-channel mismatches in high-speed
  high-resolution photonic analog-to-digital converter},\ }\href
  {https://doi.org/10.1364/OE.24.024061} {\bibfield  {journal} {\bibinfo
  {journal} {Opt. Express}\ }\textbf {\bibinfo {volume} {24}},\ \bibinfo
  {pages} {24061} (\bibinfo {year} {2016})}\BibitemShut {NoStop}%
\bibitem [{\citenamefont {Xu}\ \emph {et~al.}(2015)\citenamefont {Xu},
  \citenamefont {Zheng}, \citenamefont {Chen}, \citenamefont {Chi},
  \citenamefont {Jin},\ and\ \citenamefont {Zhang}}]{Xu_2016}%
  \BibitemOpen
  \bibfield  {author} {\bibinfo {author} {\bibfnamefont {C.}~\bibnamefont
  {Xu}}, \bibinfo {author} {\bibfnamefont {S.}~\bibnamefont {Zheng}}, \bibinfo
  {author} {\bibfnamefont {X.}~\bibnamefont {Chen}}, \bibinfo {author}
  {\bibfnamefont {H.}~\bibnamefont {Chi}}, \bibinfo {author} {\bibfnamefont
  {X.}~\bibnamefont {Jin}},\ and\ \bibinfo {author} {\bibfnamefont
  {X.}~\bibnamefont {Zhang}},\ }\bibfield  {title} {\bibinfo {title}
  {Photonic-assisted time-interleaved adc based on optical delay line},\ }\href
  {https://doi.org/10.1088/2040-8978/18/1/015704} {\bibfield  {journal}
  {\bibinfo  {journal} {Journal of Optics}\ }\textbf {\bibinfo {volume} {18}},\
  \bibinfo {pages} {015704} (\bibinfo {year} {2015})}\BibitemShut {NoStop}%
\bibitem [{\citenamefont {Wang}\ \emph {et~al.}(2023)\citenamefont {Wang},
  \citenamefont {Zhou}, \citenamefont {Min}, \citenamefont {Du},\ and\
  \citenamefont {Wang}}]{photonics10070817}%
  \BibitemOpen
  \bibfield  {author} {\bibinfo {author} {\bibfnamefont {G.}~\bibnamefont
  {Wang}}, \bibinfo {author} {\bibfnamefont {Y.}~\bibnamefont {Zhou}}, \bibinfo
  {author} {\bibfnamefont {R.}~\bibnamefont {Min}}, \bibinfo {author}
  {\bibfnamefont {E.}~\bibnamefont {Du}},\ and\ \bibinfo {author}
  {\bibfnamefont {C.}~\bibnamefont {Wang}},\ }\bibfield  {title} {\bibinfo
  {title} {Principle and recent development in photonic time-stretch imaging},\
  }\bibfield  {journal} {\bibinfo  {journal} {Photonics}\ }\textbf {\bibinfo
  {volume} {10}},\ \href {https://doi.org/10.3390/photonics10070817}
  {10.3390/photonics10070817} (\bibinfo {year} {2023})\BibitemShut {NoStop}%
\bibitem [{\citenamefont {Godin}\ \emph {et~al.}(2022)\citenamefont {Godin},
  \citenamefont {Sader}, \citenamefont {Kashi}, \citenamefont {Hanzard},
  \citenamefont {Hideur}, \citenamefont {Moss}, \citenamefont {Morandotti},
  \citenamefont {Genty}, \citenamefont {Dudley}, \citenamefont {Pasquazi},
  \citenamefont {Kues},\ and\ \citenamefont {Wetzel}}]{Godin31122022}%
  \BibitemOpen
  \bibfield  {author} {\bibinfo {author} {\bibfnamefont {T.}~\bibnamefont
  {Godin}}, \bibinfo {author} {\bibfnamefont {L.}~\bibnamefont {Sader}},
  \bibinfo {author} {\bibfnamefont {A.~K.}\ \bibnamefont {Kashi}}, \bibinfo
  {author} {\bibfnamefont {P.-H.}\ \bibnamefont {Hanzard}}, \bibinfo {author}
  {\bibfnamefont {A.}~\bibnamefont {Hideur}}, \bibinfo {author} {\bibfnamefont
  {D.~J.}\ \bibnamefont {Moss}}, \bibinfo {author} {\bibfnamefont
  {R.}~\bibnamefont {Morandotti}}, \bibinfo {author} {\bibfnamefont
  {G.}~\bibnamefont {Genty}}, \bibinfo {author} {\bibfnamefont {J.~M.}\
  \bibnamefont {Dudley}}, \bibinfo {author} {\bibfnamefont {A.}~\bibnamefont
  {Pasquazi}}, \bibinfo {author} {\bibfnamefont {M.}~\bibnamefont {Kues}},\
  and\ \bibinfo {author} {\bibfnamefont {B.}~\bibnamefont {Wetzel}},\
  }\bibfield  {title} {\bibinfo {title} {Recent advances on time-stretch
  dispersive fourier transform and its applications},\ }\href
  {https://doi.org/10.1080/23746149.2022.2067487} {\bibfield  {journal}
  {\bibinfo  {journal} {Advances in Physics: X}\ }\textbf {\bibinfo {volume}
  {7}},\ \bibinfo {pages} {2067487} (\bibinfo {year} {2022})},\ \Eprint
  {https://arxiv.org/abs/https://doi.org/10.1080/23746149.2022.2067487}
  {https://doi.org/10.1080/23746149.2022.2067487} \BibitemShut {NoStop}%
\bibitem [{\citenamefont {Lei}\ \emph {et~al.}()\citenamefont {Lei},
  \citenamefont {Guo}, \citenamefont {Cheng},\ and\ \citenamefont
  {Goda}}]{10.1063/1.4941050}%
  \BibitemOpen
  \bibfield  {author} {\bibinfo {author} {\bibfnamefont {C.}~\bibnamefont
  {Lei}}, \bibinfo {author} {\bibfnamefont {B.}~\bibnamefont {Guo}}, \bibinfo
  {author} {\bibfnamefont {Z.}~\bibnamefont {Cheng}},\ and\ \bibinfo {author}
  {\bibfnamefont {K.}~\bibnamefont {Goda}}\ }\href
  {https://doi.org/10.1063/1.4941050} {10.1063/1.4941050}\BibitemShut {NoStop}%
\bibitem [{\citenamefont {Mahjoubfar}\ \emph {et~al.}(2017)\citenamefont
  {Mahjoubfar}, \citenamefont {Churkin}, \citenamefont {Barland}, \citenamefont
  {Broderick}, \citenamefont {Turitsyn},\ and\ \citenamefont
  {Jalali}}]{Mahjoubfar:2017aa}%
  \BibitemOpen
  \bibfield  {author} {\bibinfo {author} {\bibfnamefont {A.}~\bibnamefont
  {Mahjoubfar}}, \bibinfo {author} {\bibfnamefont {D.~V.}\ \bibnamefont
  {Churkin}}, \bibinfo {author} {\bibfnamefont {S.}~\bibnamefont {Barland}},
  \bibinfo {author} {\bibfnamefont {N.}~\bibnamefont {Broderick}}, \bibinfo
  {author} {\bibfnamefont {S.~K.}\ \bibnamefont {Turitsyn}},\ and\ \bibinfo
  {author} {\bibfnamefont {B.}~\bibnamefont {Jalali}},\ }\bibfield  {title}
  {\bibinfo {title} {Time stretch and its applications},\ }\href
  {https://doi.org/10.1038/nphoton.2017.76} {\bibfield  {journal} {\bibinfo
  {journal} {Nature Photonics}\ }\textbf {\bibinfo {volume} {11}},\ \bibinfo
  {pages} {341} (\bibinfo {year} {2017})}\BibitemShut {NoStop}%
\bibitem [{\citenamefont {Coppinger}\ \emph {et~al.}(1999)\citenamefont
  {Coppinger}, \citenamefont {Bhushan},\ and\ \citenamefont {Jalali}}]{775471}%
  \BibitemOpen
  \bibfield  {author} {\bibinfo {author} {\bibfnamefont {F.}~\bibnamefont
  {Coppinger}}, \bibinfo {author} {\bibfnamefont {A.}~\bibnamefont {Bhushan}},\
  and\ \bibinfo {author} {\bibfnamefont {B.}~\bibnamefont {Jalali}},\
  }\bibfield  {title} {\bibinfo {title} {Photonic time stretch and its
  application to analog-to-digital conversion},\ }\href
  {https://doi.org/10.1109/22.775471} {\bibfield  {journal} {\bibinfo
  {journal} {IEEE Transactions on Microwave Theory and Techniques}\ }\textbf
  {\bibinfo {volume} {47}},\ \bibinfo {pages} {1309} (\bibinfo {year}
  {1999})}\BibitemShut {NoStop}%
\bibitem [{\citenamefont {Zhou}\ \emph {et~al.}(2022)\citenamefont {Zhou},
  \citenamefont {Li},\ and\ \citenamefont {Chi}}]{Zhou:22}%
  \BibitemOpen
  \bibfield  {author} {\bibinfo {author} {\bibfnamefont {H.}~\bibnamefont
  {Zhou}}, \bibinfo {author} {\bibfnamefont {R.}~\bibnamefont {Li}},\ and\
  \bibinfo {author} {\bibfnamefont {H.}~\bibnamefont {Chi}},\ }\bibfield
  {title} {\bibinfo {title} {Optical signal sampling based on compressive
  sensing with adjustable compression ratio},\ }\href
  {https://opg.optica.org/copp/abstract.cfm?URI=copp-6-3-288} {\bibfield
  {journal} {\bibinfo  {journal} {Curr. Opt. Photon.}\ }\textbf {\bibinfo
  {volume} {6}},\ \bibinfo {pages} {288} (\bibinfo {year} {2022})}\BibitemShut
  {NoStop}%
\bibitem [{\citenamefont {Yang}\ \emph {et~al.}(2023)\citenamefont {Yang},
  \citenamefont {Xu}, \citenamefont {Chi}, \citenamefont {Liu},\ and\
  \citenamefont {Yang}}]{Yang:23}%
  \BibitemOpen
  \bibfield  {author} {\bibinfo {author} {\bibfnamefont {B.}~\bibnamefont
  {Yang}}, \bibinfo {author} {\bibfnamefont {Q.}~\bibnamefont {Xu}}, \bibinfo
  {author} {\bibfnamefont {H.}~\bibnamefont {Chi}}, \bibinfo {author}
  {\bibfnamefont {Z.}~\bibnamefont {Liu}},\ and\ \bibinfo {author}
  {\bibfnamefont {S.}~\bibnamefont {Yang}},\ }\bibfield  {title} {\bibinfo
  {title} {Photonic compressive sampling of wideband sparse radio frequency
  signals with 1-bit quantization},\ }\href {https://doi.org/10.1364/OE.486976}
  {\bibfield  {journal} {\bibinfo  {journal} {Opt. Express}\ }\textbf {\bibinfo
  {volume} {31}},\ \bibinfo {pages} {18159} (\bibinfo {year}
  {2023})}\BibitemShut {NoStop}%
\bibitem [{\citenamefont {Jaeger}\ and\ \citenamefont {Haas}()}]{Jaeger78}%
  \BibitemOpen
  \bibfield  {author} {\bibinfo {author} {\bibfnamefont {H.}~\bibnamefont
  {Jaeger}}\ and\ \bibinfo {author} {\bibfnamefont {H.}~\bibnamefont {Haas}}\
  }\href {https://doi.org/10.1126/science.1091277}
  {10.1126/science.1091277}\BibitemShut {NoStop}%
\bibitem [{\citenamefont {Tanaka}\ \emph {et~al.}(2019)\citenamefont {Tanaka},
  \citenamefont {Yamane}, \citenamefont {H{\'e}roux}, \citenamefont {Nakane},
  \citenamefont {Kanazawa}, \citenamefont {Takeda}, \citenamefont {Numata},
  \citenamefont {Nakano},\ and\ \citenamefont {Hirose}}]{Tanaka2019}%
  \BibitemOpen
  \bibfield  {author} {\bibinfo {author} {\bibfnamefont {G.}~\bibnamefont
  {Tanaka}}, \bibinfo {author} {\bibfnamefont {T.}~\bibnamefont {Yamane}},
  \bibinfo {author} {\bibfnamefont {J.~B.}\ \bibnamefont {H{\'e}roux}},
  \bibinfo {author} {\bibfnamefont {R.}~\bibnamefont {Nakane}}, \bibinfo
  {author} {\bibfnamefont {N.}~\bibnamefont {Kanazawa}}, \bibinfo {author}
  {\bibfnamefont {S.}~\bibnamefont {Takeda}}, \bibinfo {author} {\bibfnamefont
  {H.}~\bibnamefont {Numata}}, \bibinfo {author} {\bibfnamefont
  {D.}~\bibnamefont {Nakano}},\ and\ \bibinfo {author} {\bibfnamefont
  {A.}~\bibnamefont {Hirose}},\ }\bibfield  {title} {\bibinfo {title} {Recent
  advances in physical reservoir computing: A review},\ }\href@noop {}
  {\bibfield  {journal} {\bibinfo  {journal} {Neural Networks}\ }\textbf
  {\bibinfo {volume} {115}},\ \bibinfo {pages} {100} (\bibinfo {year}
  {2019})}\BibitemShut {NoStop}%
\bibitem [{\citenamefont {Yan}\ \emph {et~al.}(2024)\citenamefont {Yan},
  \citenamefont {Huang}, \citenamefont {Bienstman}, \citenamefont {Tino},
  \citenamefont {Lin},\ and\ \citenamefont {Sun}}]{Yan:2024aa}%
  \BibitemOpen
  \bibfield  {author} {\bibinfo {author} {\bibfnamefont {M.}~\bibnamefont
  {Yan}}, \bibinfo {author} {\bibfnamefont {C.}~\bibnamefont {Huang}}, \bibinfo
  {author} {\bibfnamefont {P.}~\bibnamefont {Bienstman}}, \bibinfo {author}
  {\bibfnamefont {P.}~\bibnamefont {Tino}}, \bibinfo {author} {\bibfnamefont
  {W.}~\bibnamefont {Lin}},\ and\ \bibinfo {author} {\bibfnamefont
  {J.}~\bibnamefont {Sun}},\ }\bibfield  {title} {\bibinfo {title} {Emerging
  opportunities and challenges for the future of reservoir computing},\ }\href
  {https://doi.org/10.1038/s41467-024-45187-1} {\bibfield  {journal} {\bibinfo
  {journal} {Nature Communications}\ }\textbf {\bibinfo {volume} {15}},\
  \bibinfo {pages} {2056} (\bibinfo {year} {2024})}\BibitemShut {NoStop}%
\bibitem [{\citenamefont {der Sande}\ \emph {et~al.}(2017)\citenamefont {der
  Sande}, \citenamefont {Brunner},\ and\ \citenamefont
  {Soriano}}]{VanderSandeBrunnerSoriano+2017+561+576}%
  \BibitemOpen
  \bibfield  {author} {\bibinfo {author} {\bibfnamefont {G.~V.}\ \bibnamefont
  {der Sande}}, \bibinfo {author} {\bibfnamefont {D.}~\bibnamefont {Brunner}},\
  and\ \bibinfo {author} {\bibfnamefont {M.~C.}\ \bibnamefont {Soriano}},\
  }\bibfield  {title} {\bibinfo {title} {Advances in photonic reservoir
  computing},\ }\href {https://doi.org/doi:10.1515/nanoph-2016-0132} {\bibfield
   {journal} {\bibinfo  {journal} {Nanophotonics}\ }\textbf {\bibinfo {volume}
  {6}},\ \bibinfo {pages} {561} (\bibinfo {year} {2017})}\BibitemShut {NoStop}%
\bibitem [{\citenamefont {Brunner}\ \emph {et~al.}(2013)\citenamefont
  {Brunner}, \citenamefont {Soriano}, \citenamefont {Mirasso},\ and\
  \citenamefont {Fischer}}]{Brunner2013}%
  \BibitemOpen
  \bibfield  {author} {\bibinfo {author} {\bibfnamefont {D.}~\bibnamefont
  {Brunner}}, \bibinfo {author} {\bibfnamefont {M.~C.}\ \bibnamefont
  {Soriano}}, \bibinfo {author} {\bibfnamefont {C.~R.}\ \bibnamefont
  {Mirasso}},\ and\ \bibinfo {author} {\bibfnamefont {I.}~\bibnamefont
  {Fischer}},\ }\bibfield  {title} {\bibinfo {title} {Parallel photonic
  information processing at gigabyte per second data rates using transient
  states},\ }\href@noop {} {\bibfield  {journal} {\bibinfo  {journal} {Nature
  Communications}\ }\textbf {\bibinfo {volume} {4}},\ \bibinfo {pages} {1364}
  (\bibinfo {year} {2013})}\BibitemShut {NoStop}%
\bibitem [{\citenamefont {Takano}\ \emph {et~al.}(2018)\citenamefont {Takano},
  \citenamefont {Sugano}, \citenamefont {Inubushi}, \citenamefont {Yoshimura},
  \citenamefont {Sunada}, \citenamefont {Kanno},\ and\ \citenamefont
  {Uchida}}]{Takano2018}%
  \BibitemOpen
  \bibfield  {author} {\bibinfo {author} {\bibfnamefont {K.}~\bibnamefont
  {Takano}}, \bibinfo {author} {\bibfnamefont {C.}~\bibnamefont {Sugano}},
  \bibinfo {author} {\bibfnamefont {M.}~\bibnamefont {Inubushi}}, \bibinfo
  {author} {\bibfnamefont {K.}~\bibnamefont {Yoshimura}}, \bibinfo {author}
  {\bibfnamefont {S.}~\bibnamefont {Sunada}}, \bibinfo {author} {\bibfnamefont
  {K.}~\bibnamefont {Kanno}},\ and\ \bibinfo {author} {\bibfnamefont
  {A.}~\bibnamefont {Uchida}},\ }\bibfield  {title} {\bibinfo {title} {Compact
  reservoir computing with a photonic integrated circuit},\ }\href@noop {}
  {\bibfield  {journal} {\bibinfo  {journal} {Optics express}\ }\textbf
  {\bibinfo {volume} {26}},\ \bibinfo {pages} {29424} (\bibinfo {year}
  {2018})}\BibitemShut {NoStop}%
\bibitem [{\citenamefont {Vandoorne}\ \emph {et~al.}(2014)\citenamefont
  {Vandoorne}, \citenamefont {Mechet}, \citenamefont {Van~Vaerenbergh},
  \citenamefont {Fiers}, \citenamefont {Morthier}, \citenamefont {Verstraeten},
  \citenamefont {Schrauwen}, \citenamefont {Dambre},\ and\ \citenamefont
  {Bienstman}}]{Vandoorne2014}%
  \BibitemOpen
  \bibfield  {author} {\bibinfo {author} {\bibfnamefont {K.}~\bibnamefont
  {Vandoorne}}, \bibinfo {author} {\bibfnamefont {P.}~\bibnamefont {Mechet}},
  \bibinfo {author} {\bibfnamefont {T.}~\bibnamefont {Van~Vaerenbergh}},
  \bibinfo {author} {\bibfnamefont {M.}~\bibnamefont {Fiers}}, \bibinfo
  {author} {\bibfnamefont {G.}~\bibnamefont {Morthier}}, \bibinfo {author}
  {\bibfnamefont {D.}~\bibnamefont {Verstraeten}}, \bibinfo {author}
  {\bibfnamefont {B.}~\bibnamefont {Schrauwen}}, \bibinfo {author}
  {\bibfnamefont {J.}~\bibnamefont {Dambre}},\ and\ \bibinfo {author}
  {\bibfnamefont {P.}~\bibnamefont {Bienstman}},\ }\bibfield  {title} {\bibinfo
  {title} {Experimental demonstration of reservoir computing on a silicon
  photonics chip},\ }\href@noop {} {\bibfield  {journal} {\bibinfo  {journal}
  {Nature Communications}\ }\textbf {\bibinfo {volume} {5}},\ \bibinfo {pages}
  {3541} (\bibinfo {year} {2014})}\BibitemShut {NoStop}%
\bibitem [{\citenamefont {Wang}\ \emph {et~al.}(2025)\citenamefont {Wang},
  \citenamefont {Hu}, \citenamefont {Baek}, \citenamefont {Tsuchiyama},
  \citenamefont {Joly}, \citenamefont {Liu},\ and\ \citenamefont
  {Gigan}}]{Wang2025}%
  \BibitemOpen
  \bibfield  {author} {\bibinfo {author} {\bibfnamefont {H.}~\bibnamefont
  {Wang}}, \bibinfo {author} {\bibfnamefont {J.}~\bibnamefont {Hu}}, \bibinfo
  {author} {\bibfnamefont {Y.}~\bibnamefont {Baek}}, \bibinfo {author}
  {\bibfnamefont {K.}~\bibnamefont {Tsuchiyama}}, \bibinfo {author}
  {\bibfnamefont {M.}~\bibnamefont {Joly}}, \bibinfo {author} {\bibfnamefont
  {Q.}~\bibnamefont {Liu}},\ and\ \bibinfo {author} {\bibfnamefont
  {S.}~\bibnamefont {Gigan}},\ }\bibfield  {title} {\bibinfo {title} {Optical
  next generation reservoir computing},\ }\href@noop {} {\bibfield  {journal}
  {\bibinfo  {journal} {Light: Science \& Applications}\ }\textbf {\bibinfo
  {volume} {14}},\ \bibinfo {pages} {245} (\bibinfo {year} {2025})}\BibitemShut
  {NoStop}%
\bibitem [{\citenamefont {Wang}\ \emph {et~al.}(2024)\citenamefont {Wang},
  \citenamefont {Nie}, \citenamefont {Hu}, \citenamefont {Tsang},\ and\
  \citenamefont {Huang}}]{DWang2024}%
  \BibitemOpen
  \bibfield  {author} {\bibinfo {author} {\bibfnamefont {D.}~\bibnamefont
  {Wang}}, \bibinfo {author} {\bibfnamefont {Y.}~\bibnamefont {Nie}}, \bibinfo
  {author} {\bibfnamefont {G.}~\bibnamefont {Hu}}, \bibinfo {author}
  {\bibfnamefont {H.~K.}\ \bibnamefont {Tsang}},\ and\ \bibinfo {author}
  {\bibfnamefont {C.}~\bibnamefont {Huang}},\ }\bibfield  {title} {\bibinfo
  {title} {Ultrafast silicon photonic reservoir computing engine delivering
  over 200 tops},\ }\href@noop {} {\bibfield  {journal} {\bibinfo  {journal}
  {Nature Communications}\ }\textbf {\bibinfo {volume} {15}},\ \bibinfo {pages}
  {10841} (\bibinfo {year} {2024})}\BibitemShut {NoStop}%
\bibitem [{\citenamefont {Sunada}\ and\ \citenamefont
  {Uchida}(2021{\natexlab{a}})}]{Sunada2021}%
  \BibitemOpen
  \bibfield  {author} {\bibinfo {author} {\bibfnamefont {S.}~\bibnamefont
  {Sunada}}\ and\ \bibinfo {author} {\bibfnamefont {A.}~\bibnamefont
  {Uchida}},\ }\bibfield  {title} {\bibinfo {title} {Photonic neural field on a
  silicon chip: large-scale, high-speed neuro-inspired computing and sensing},\
  }\href@noop {} {\bibfield  {journal} {\bibinfo  {journal} {Optica}\ }\textbf
  {\bibinfo {volume} {8}},\ \bibinfo {pages} {1388} (\bibinfo {year}
  {2021}{\natexlab{a}})}\BibitemShut {NoStop}%
\bibitem [{\citenamefont {Roussel}\ \emph {et~al.}(2022)\citenamefont
  {Roussel}, \citenamefont {Szwaj}, \citenamefont {Evain}, \citenamefont
  {Steffen}, \citenamefont {Gerth}, \citenamefont {Jalali},\ and\ \citenamefont
  {Bielawski}}]{Roussel:2022aa}%
  \BibitemOpen
  \bibfield  {author} {\bibinfo {author} {\bibfnamefont {E.}~\bibnamefont
  {Roussel}}, \bibinfo {author} {\bibfnamefont {C.}~\bibnamefont {Szwaj}},
  \bibinfo {author} {\bibfnamefont {C.}~\bibnamefont {Evain}}, \bibinfo
  {author} {\bibfnamefont {B.}~\bibnamefont {Steffen}}, \bibinfo {author}
  {\bibfnamefont {C.}~\bibnamefont {Gerth}}, \bibinfo {author} {\bibfnamefont
  {B.}~\bibnamefont {Jalali}},\ and\ \bibinfo {author} {\bibfnamefont
  {S.}~\bibnamefont {Bielawski}},\ }\bibfield  {title} {\bibinfo {title} {Phase
  diversity electro-optic sampling: A new approach to single-shot terahertz
  waveform recording},\ }\href {https://doi.org/10.1038/s41377-021-00696-2}
  {\bibfield  {journal} {\bibinfo  {journal} {Light: Science \& Applications}\
  }\textbf {\bibinfo {volume} {11}},\ \bibinfo {pages} {14} (\bibinfo {year}
  {2022})}\BibitemShut {NoStop}%
\bibitem [{\citenamefont {Tikan}\ \emph {et~al.}(2018)\citenamefont {Tikan},
  \citenamefont {Bielawski}, \citenamefont {Szwaj}, \citenamefont {Randoux},\
  and\ \citenamefont {Suret}}]{Tikan:2018aa}%
  \BibitemOpen
  \bibfield  {author} {\bibinfo {author} {\bibfnamefont {A.}~\bibnamefont
  {Tikan}}, \bibinfo {author} {\bibfnamefont {S.}~\bibnamefont {Bielawski}},
  \bibinfo {author} {\bibfnamefont {C.}~\bibnamefont {Szwaj}}, \bibinfo
  {author} {\bibfnamefont {S.}~\bibnamefont {Randoux}},\ and\ \bibinfo {author}
  {\bibfnamefont {P.}~\bibnamefont {Suret}},\ }\bibfield  {title} {\bibinfo
  {title} {Single-shot measurement of phase and amplitude by using a heterodyne
  time-lens system and ultrafast digital time-holography},\ }\href
  {https://doi.org/10.1038/s41566-018-0113-8} {\bibfield  {journal} {\bibinfo
  {journal} {Nature Photonics}\ }\textbf {\bibinfo {volume} {12}},\ \bibinfo
  {pages} {228} (\bibinfo {year} {2018})}\BibitemShut {NoStop}%
\bibitem [{\citenamefont {Suret}\ \emph {et~al.}(2016)\citenamefont {Suret},
  \citenamefont {Koussaifi}, \citenamefont {Tikan}, \citenamefont {Evain},
  \citenamefont {Randoux}, \citenamefont {Szwaj},\ and\ \citenamefont
  {Bielawski}}]{Suret:2016aa}%
  \BibitemOpen
  \bibfield  {author} {\bibinfo {author} {\bibfnamefont {P.}~\bibnamefont
  {Suret}}, \bibinfo {author} {\bibfnamefont {R.~E.}\ \bibnamefont
  {Koussaifi}}, \bibinfo {author} {\bibfnamefont {A.}~\bibnamefont {Tikan}},
  \bibinfo {author} {\bibfnamefont {C.}~\bibnamefont {Evain}}, \bibinfo
  {author} {\bibfnamefont {S.}~\bibnamefont {Randoux}}, \bibinfo {author}
  {\bibfnamefont {C.}~\bibnamefont {Szwaj}},\ and\ \bibinfo {author}
  {\bibfnamefont {S.}~\bibnamefont {Bielawski}},\ }\bibfield  {title} {\bibinfo
  {title} {Single-shot observation of optical rogue waves in integrable
  turbulence using time microscopy},\ }\href
  {https://doi.org/10.1038/ncomms13136} {\bibfield  {journal} {\bibinfo
  {journal} {Nature Communications}\ }\textbf {\bibinfo {volume} {7}},\
  \bibinfo {pages} {13136} (\bibinfo {year} {2016})}\BibitemShut {NoStop}%
\bibitem [{\citenamefont {Sunada}\ and\ \citenamefont
  {Uchida}(2021{\natexlab{b}})}]{Sunada:21}%
  \BibitemOpen
  \bibfield  {author} {\bibinfo {author} {\bibfnamefont {S.}~\bibnamefont
  {Sunada}}\ and\ \bibinfo {author} {\bibfnamefont {A.}~\bibnamefont
  {Uchida}},\ }\bibfield  {title} {\bibinfo {title} {Photonic neural field on a
  silicon chip: large-scale, high-speed neuro-inspired computing and sensing},\
  }\href {https://doi.org/10.1364/OPTICA.434918} {\bibfield  {journal}
  {\bibinfo  {journal} {Optica}\ }\textbf {\bibinfo {volume} {8}},\ \bibinfo
  {pages} {1388} (\bibinfo {year} {2021}{\natexlab{b}})}\BibitemShut {NoStop}%
\bibitem [{\citenamefont {Jaeger}(2001)}]{article}%
  \BibitemOpen
  \bibfield  {author} {\bibinfo {author} {\bibfnamefont {H.}~\bibnamefont
  {Jaeger}},\ }\bibfield  {title} {\bibinfo {title} {The" echo state" approach
  to analysing and training recurrent neural networks-with an erratum note'},\
  }\href@noop {} {\bibfield  {journal} {\bibinfo  {journal} {Bonn, Germany:
  German National Research Center for Information Technology GMD Technical
  Report}\ }\textbf {\bibinfo {volume} {148}} (\bibinfo {year}
  {2001})}\BibitemShut {NoStop}%
\bibitem [{\citenamefont {Bunimovich}(1974)}]{Bunimovich:1974aa}%
  \BibitemOpen
  \bibfield  {author} {\bibinfo {author} {\bibfnamefont {L.~A.}\ \bibnamefont
  {Bunimovich}},\ }\bibfield  {title} {\bibinfo {title} {On ergodic properties
  of certain billiards},\ }\href {https://doi.org/10.1007/BF01075700}
  {\bibfield  {journal} {\bibinfo  {journal} {Functional Analysis and Its
  Applications}\ }\textbf {\bibinfo {volume} {8}},\ \bibinfo {pages} {254}
  (\bibinfo {year} {1974})}\BibitemShut {NoStop}%
\bibitem [{\citenamefont {Yamaguchi}\ \emph {et~al.}(2023)\citenamefont
  {Yamaguchi}, \citenamefont {Arai}, \citenamefont {Niiyama}, \citenamefont
  {Uchida},\ and\ \citenamefont {Sunada}}]{Yamaguchi:2023aa}%
  \BibitemOpen
  \bibfield  {author} {\bibinfo {author} {\bibfnamefont {T.}~\bibnamefont
  {Yamaguchi}}, \bibinfo {author} {\bibfnamefont {K.}~\bibnamefont {Arai}},
  \bibinfo {author} {\bibfnamefont {T.}~\bibnamefont {Niiyama}}, \bibinfo
  {author} {\bibfnamefont {A.}~\bibnamefont {Uchida}},\ and\ \bibinfo {author}
  {\bibfnamefont {S.}~\bibnamefont {Sunada}},\ }\bibfield  {title} {\bibinfo
  {title} {Time-domain photonic image processor based on speckle projection and
  reservoir computing},\ }\href {https://doi.org/10.1038/s42005-023-01368-w}
  {\bibfield  {journal} {\bibinfo  {journal} {Communications Physics}\ }\textbf
  {\bibinfo {volume} {6}},\ \bibinfo {pages} {250} (\bibinfo {year}
  {2023})}\BibitemShut {NoStop}%
\bibitem [{\citenamefont {You}\ \emph {et~al.}(2025)\citenamefont {You},
  \citenamefont {Arai},\ and\ \citenamefont {Sunada}}]{You:25}%
  \BibitemOpen
  \bibfield  {author} {\bibinfo {author} {\bibfnamefont {M.}~\bibnamefont
  {You}}, \bibinfo {author} {\bibfnamefont {K.}~\bibnamefont {Arai}},\ and\
  \bibinfo {author} {\bibfnamefont {S.}~\bibnamefont {Sunada}},\ }\bibfield
  {title} {\bibinfo {title} {Nonlinear time series computing using a linear
  optical microcavity},\ }\href {https://doi.org/10.1364/OE.559262} {\bibfield
  {journal} {\bibinfo  {journal} {Opt. Express}\ }\textbf {\bibinfo {volume}
  {33}},\ \bibinfo {pages} {24982} (\bibinfo {year} {2025})}\BibitemShut
  {NoStop}%
\bibitem [{\citenamefont {Weigend}\ and\ \citenamefont
  {Gershenfeld}(1993)}]{Weigend1993}%
  \BibitemOpen
  \bibfield  {author} {\bibinfo {author} {\bibfnamefont {A.~S.}\ \bibnamefont
  {Weigend}}\ and\ \bibinfo {author} {\bibfnamefont {N.~A.}\ \bibnamefont
  {Gershenfeld}},\ }\bibfield  {title} {\bibinfo {title} {Results of the time
  series prediction competition at the santa fe institute},\ }in\ \href@noop {}
  {\emph {\bibinfo {booktitle} {IEEE international conference on neural
  networks}}}\ (\bibinfo {organization} {IEEE},\ \bibinfo {year} {1993})\ pp.\
  \bibinfo {pages} {1786--1793}\BibitemShut {NoStop}%
\end{thebibliography}

\end{document}


\title{Expanding detection bandwidth via a photonic reservoir for ultrafast optical sensing: supplemental document}

%
%

\maketitle

\section{Waveforms used for training}
The $l$-th waveform $u_l(t)$ for training was set as follows:
\begin{align}
u_l(t) = \sum_{m = 1}^{M}\cos\left( m\Delta \omega t + \phi^{(l)}_m \right), 
\end{align} 
where $\Delta \omega$ denotes the frequency resolution, and $\phi_m^{(l)}$ were randomly sampled from uniform distributions $\phi_m^{(l)} \sim U(0,2\pi)$. $M$ was set as $B_{\rm sig}/\Delta\omega$.

\section{Reconstructed waveforms and their power spectra}
\begin{figure}[htbp]
\centering
\includegraphics[bb=0 0 398 533, width=14cm]{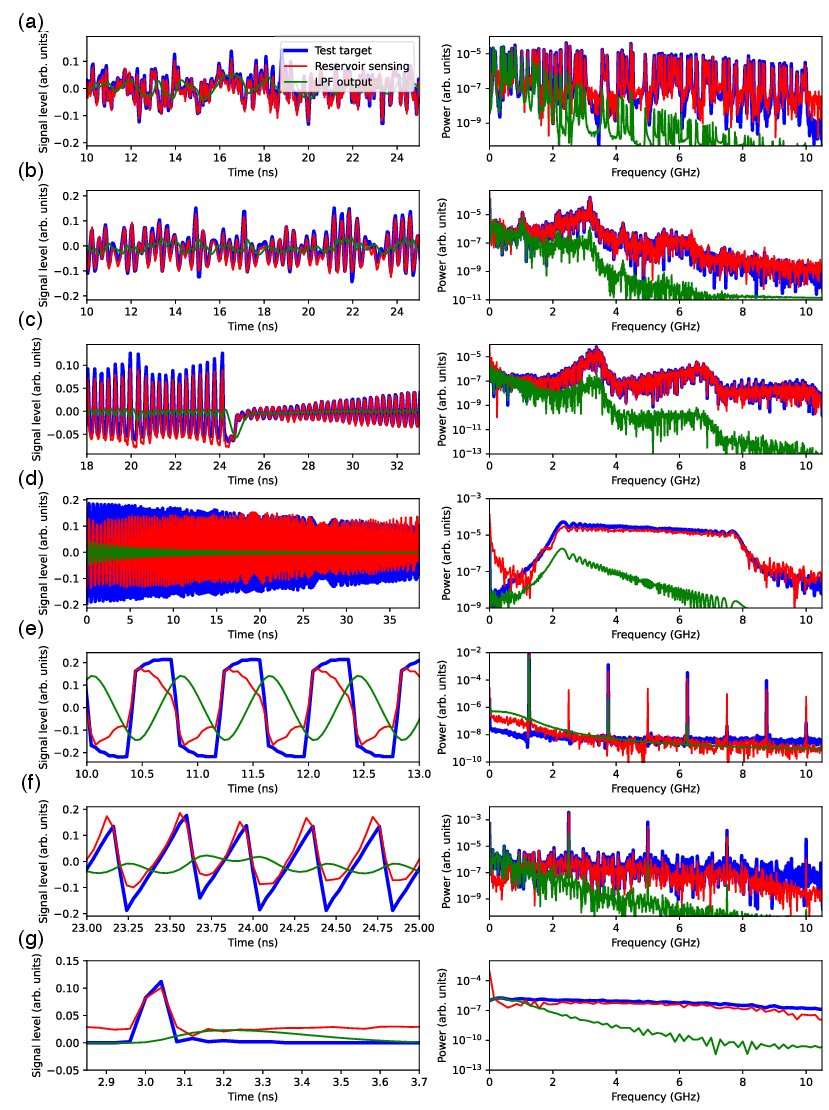}
\caption{
Reconstructed waveforms (left column) and their corresponding power spectra (right column). The detector bandwidth is $B_{\rm sensor} \approx 1 $ GHz. (a) A random waveform (with 10-GHz bandwidth), which is uncorrelated with the training samples. The reconstruction error (NMSE) was $0.06$. (b) A laser chaos signal generated based on the Lang--Kobayashi model (NMSE $=0.06$). (c) The Santa-Fe chaotic signal (NMSE $=0.09$). (d) A chirped pulse sweeping from 2 to 8~GHz over 38.4~ns (NMSE $=0.08$). (e) A periodic pulse (NMSE $=0.21$). (f) A triangular waveform (NMSE $=0.23$). (g) A single pulse with a pulse width of approximately 0.1~ns (NMSE $=0.63$). The large NMSE for the pulse signal in (e) mainly arises from a baseline (DC).}
\label{fig_all}
\end{figure}

After training, the proposed approach can reconstruct various waveforms, as discussed in Section~4.3. Figure~\ref{fig_all} presents the reconstructed waveforms, including (a) a random signal, (b) the laser chaotic time-series, (c) the Santa-Fe chaotic time-series, (d) a chirped signal, (e) a rectangular periodic waveform, (f) a triangular periodic waveform, and (g) a nearly 100-ps pulse. Their corresponding power spectra are shown in the right column in Fig.~\ref{fig_all}.

\section{Effects of reservoir nonlinearity}
\begin{figure}[htbp]
\centering
\includegraphics[bb=0 0 570 385, width=14cm]{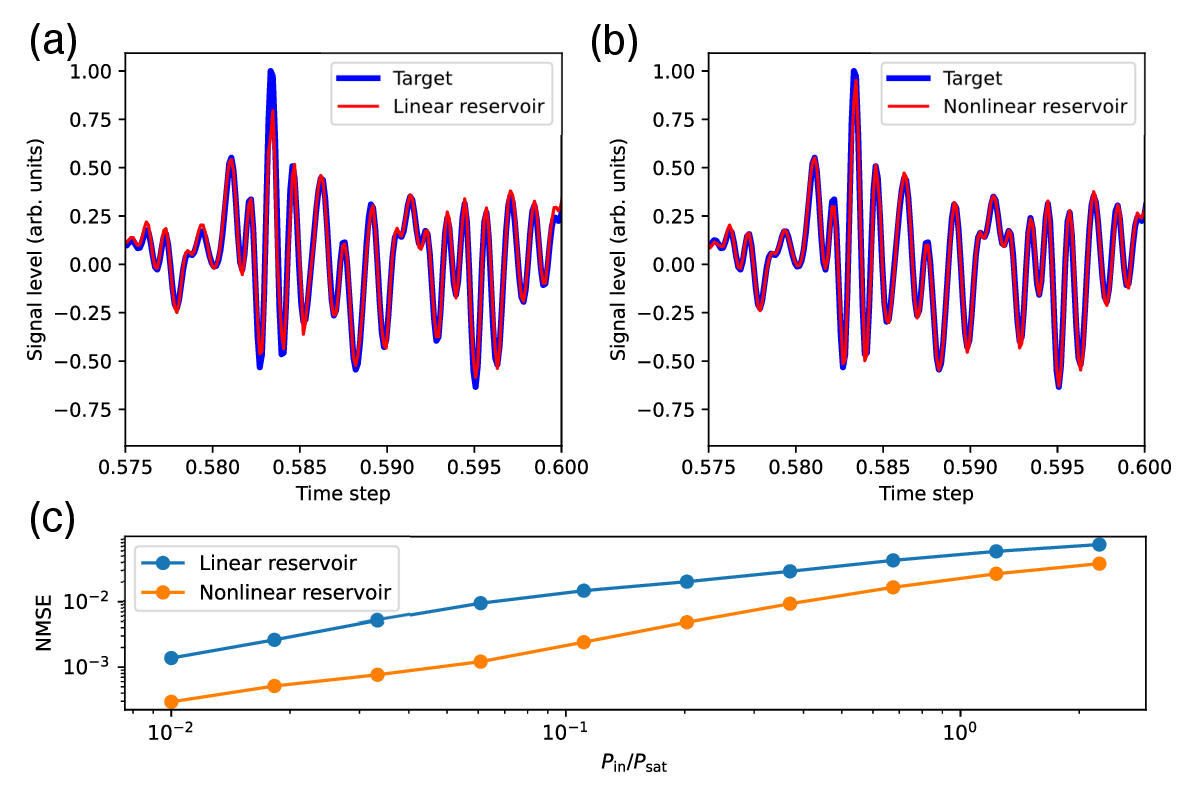}
\caption{Reconstruction results with (a) linear and (b) nonlinear reservoirs. (c) NMSE as a function of $P_{\rm in}/P_{\rm sat}$ for both reservoirs.}
\label{fig_nonlinear}
\end{figure}

We assumed that the photodetector has a nonlinear saturation, represented as $P_{\rm sat}$, as follows: 
\begin{align}
P_{\rm out} = \dfrac{
P_{\rm in}}{
1 + P_{\rm in}/P_{\rm sat}
}. \label{eq_nonlinear}
\end{align}
To evaluate the compensation capability for the photodetector's nonlinear distortion, we compared the reconstruction errors for linear and nonlinear reservoirs, which are modeled as echo-state network models with linear and $\tanh$ activation functions, respectively. The effect of the nonlinear distortion shown Eq.~(\ref{eq_nonlinear}) was considered as ${\bf x}(t) = {\bf x}(t)/\sqrt{1 + |{\bf x}|^2/P_{\rm sat}}$. 
Figure~\ref{fig_nonlinear} shows a comparison for $N = 30$, $B = 100$, and $P_{\rm in}/P_{\rm sat} = 0.09$. A linear reservoir (without nonlinearity) failed to completely compensate for the nonlinear distortion (Fig.~\ref{fig_nonlinear}(a)); consequently, the NMSE was 0.023. Meanwhile, a nonlinear reservoir can compensate for the nonlinear distortion with an NMSE of 0.006 (Fig.~\ref{fig_nonlinear}(b)). The NMSEs are shown as a function of $P_{\rm in}/P_{\rm sat}$ in Fig.~\ref{fig_nonlinear}(c). These results suggest the importance of the nonlinearity inherent in a nonlinear reservoir for compensating the nonlinearly distortion.

\section{Photonic integrated reservoir}
\begin{figure}[htbp]
\centering
\includegraphics[bb=0 0 367 162, width=8cm]{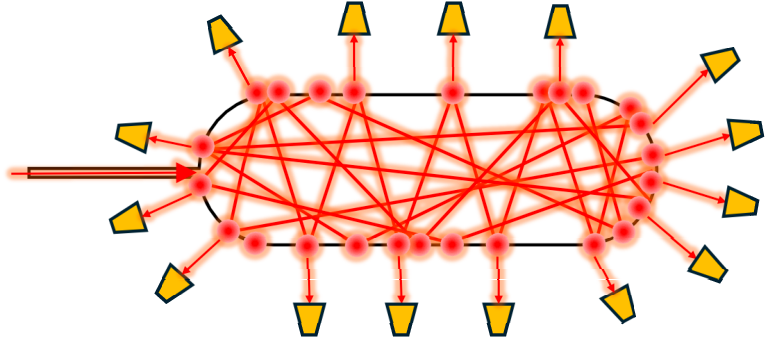}
\caption{Schematic of a photonic reservoir circuit integrated with multiple photodetectors.}
\label{fig_PD}
\end{figure}

The stadium-shaped microcavity used in this study has a large scattering loss. The leaked signals can be detected using multiple photodetectors around the stadium cavity, as shown in Fig.~\ref{fig_PD}. 

